\documentclass[aps,twocolumn,pra,tightenlines,floatfix,showpacs]{revtex4-1}
\usepackage[dvips]{graphicx}
\usepackage[english]{babel}
\usepackage{amsmath}
\usepackage{amssymb}
\usepackage{times}
\def\be{\begin{equation}}
\def\ee{\end{equation}}
\def\bea{\begin{eqnarray}}
\def\eea{\end{eqnarray}}

\renewcommand{\Re}{\mbox{Re}\,}
\def\d{\mbox{d}}
\def\i{\text{i}}
\def\e{\text{e}}
\newcommand{\ek}{\xi_{\mathbf{k}}}
\newcommand{\Ek}{{E^{}_{\mathbf{k}}}}

\newcommand{\sumk}{{\sum_{\mathbf{k}}}}

\newcommand{\sumq}{\sum_{\mathbf{q}}}

\newcommand{\Omegaq}{\Omega_{\mathbf{q}}}
\newcommand{\Gammaq}{\Gamma_{\mathbf{q}}}
\newcommand{\uk}{u_{\mathbf{k}}}
\newcommand{\vk}{v_{\mathbf{k}}}

\begin{document}

\title{Effect of the particle-hole channel on BCS--Bose-Einstein
  condensation crossover in atomic Fermi gases}

\author{Qijin Chen}

\affiliation{Department of Physics and Zhejiang Institute of Modern
  Physics, Zhejiang University, Hangzhou, Zhejiang 310027, CHINA}

\date{\today}

\begin{abstract}
  BCS--Bose-Einstein condensation (BEC) crossover is effected by
  increasing pairing strength between fermions from weak to strong in
  the particle-particle channel.  Here we study the effect of the
  particle-hole channel on the zero $T$ gap $\Delta(0)$, superfluid
  transition temperature $T_{\text{c}}$ and the pseudogap at
  $T_{\text{c}}$, as well as the mean-field ratio
  $2\Delta(0)/T_{\text{c}}^{\text{MF}}$, from BCS through BEC regimes,
  in the framework of a pairing fluctuation theory which includes
  self-consistently the contributions of finite-momentum pairs. These
  pairs necessarily lead to a pseudogap in single particle excitation
  spectrum above and below $T_{\text{c}}$. We sum over the infinite
  particle-hole ladder diagrams so that the particle-particle and
  particle-hole $T$-matrices are entangled with each other.  We find
  that the particle-hole susceptibility has a complex dynamical
  structure, with strong momentum and frequency dependencies, and is
  sensitive to temperature, gap size and interaction strength.  We
  conclude that neglecting the self-energy feedback causes a serious
  over-estimate of the particle-hole susceptibility.  In the BCS
  limit, the particle-hole channel effect may be approximated by the
  same reduction in the overall pairing strength so that the ratio
  $2\Delta(0)/T_{\text{c}}$ is unaffected, in agreement with Gor'kov
  \textit{et al.} to the leading order. However, the effect becomes
  more complex and pronounced in the crossover regime, where the
  particle-hole susceptibility is reduced by both a smaller Fermi
  surface and a big (pseudo)gap. Deep in the BEC regime, the
  particle-hole channel contributions drop to zero. We propose that
  precision measurements of the magnetic field for Feshbach resonance
  at low temperatures as a function of density can be used to quantify
  the particle-hole susceptibility and test different theories.
\end{abstract}


\maketitle

\section{Introduction}

BCS--Bose-Einstein condensation (BEC) crossover has been an
interesting research subject since 1980's
\cite{Eagles,Leggett,NSR,TDLee1,TDLee2,Serene,SadeMelo,Randeria89,Randeria92,Randeriareview,Ranninger,Micnas1,Haussmann,Tchern,Pedersen3,Uemura,Janko,Maly1,Maly2,Kosztin1,Chen2,Chen1,Chen3,Chen-Schrieffer,Strinati3,Strinati6,crossoverothers,Gyorffy,Reviews2}. The
experimental realization of BCS-BEC crossover in ultracold atomic
Fermi gases
\cite{Regal,Jin3,Jin4,Grimm,Grimm2,Grimm3,Ketterle2,Ketterle3,KetterleV,Thomas2},
with the help of Feshbach resonances, has given it a strong boost
\cite{Milstein,Griffin,Griffin2,JS2,Strinati4,Tosi,Heiselberg,Stoof,Salomon3,Jin5,Jin_us,Hulet4,ChenClosed,ChenThermo,Drummond2,Griffingroup,Grimm5,ThomasDamping,Zwierlein2006,Rice2}
over the past several years. When the pairing interaction is tuned
from weak to strong in a two component Fermi gas, the superfluid
behavior evolves continuously from the type of BCS to that of BEC
\cite{Leggett,NSR,Reviews2}.

In such a fundamentally fermionic system, superfluidity mainly
concerns pairing, namely, interactions in the particle-particle
channel. In contrast, the particle-hole channel mainly causes a
chemical potential shift, and is often neglected
\cite{Schrieffer}. For example, in a conventional superconductor, the
chemical potential below and above $T_{\text{c}}$ are essentially the
same, and thus its dependence on the temperature and the interaction
strength has been completely neglected in the weak coupling BCS theory
for normal metal superconductors. On the other hand, Gor'kov and
Melik-Barkhudarov (GMB) \cite{GMB} considered the lowest order
correction from the particle-hole channel, (which has been referred to
as induced interaction in the literature), and found that both
$T_{\text{c}}$ and zero temperature gap $\Delta(0)$ are suppressed by
a \emph{big} factor of $(4\e)^{1/3}\approx 2.22$. Berk and Schrieffer
\cite{BerkSchrieffer} also studied a similar effect in the form of
ferromagnetic spin correlations in superconductors. Despite the big
size of the GMB correction, the effect of the particle-hole channel
has been largely overlooked in the theoretical study of BCS-BEC
crossover, until it has become realistic to achieve such crossover
experimentally in atomic Fermi gases. Heiselberg and coworkers
\cite{Heiselberg2000} considered the effect of the lowest order
induced interaction in dilute Fermi gases and generalized it to the
case of multispecies of fermions as well as the possibility of
exchange of bosons. Kim \textit{et al.}  \cite{KimTorma2009}
considered the lowest order induced interactions in optical
lattices. Within the \emph{mean-field} treatment and \emph{without}
including the excitation gap in the particle and hole propagators,
these authors found the same effective overall interaction at zero $T$
and at $T_{\text{c}}$ and hence an unaffected mean-field ratio
$2\Delta(0)/k_{\text{B}}T_{\text{c}} \approx 3.53$.  Martikainen
\textit{et al.}  \cite{TormaPethick2009} considered the lowest order
induced interactions in a three-component Fermi gas. It has become
clear that including only the perturbative lowest order induced
interaction is \emph{not} appropriate away from the weak coupling BCS
regime. Yin and coworkers \cite{Yin2009} went beyond the lowest order
and considered the induced interactions from all particle-hole ladder
diagrams, i.e., the entire particle-hole $T$-matrix. However, in all
the above works, only the \emph{bare} particle-hole susceptibility
$\chi_{\text{ph}}^0$ was considered, and it was averaged on-shell and
only on the Fermi surface, with equal momenta for the particle and the
hole propagators. No self-energy feedback was included. Therefore,
there was necessarily no pseudogap in the fermion excitation spectrum
at $T_{\text{c}}$. This is basically equivalent to replacing the
particle-hole susceptibility $\chi_{\text{ph}}^0$ by an essentially
temperature independent constant, leading to a simple downshift in the
pairing interaction.


As the gap and $T_{\text{c}}$ increase with interaction strength, it
can naturally be expected that the contribution from the particle-hole
channel, or the induced interaction, will acquire a significant
temperature dependence. More importantly, \emph{the presence of the
  (pseudo)gap serves to suppress the particle-hole fluctuations}
(which tend to break pairing). In other words, neglecting the feedback
of the gap related self energy in the particle-hole susceptibility is
expected to cause an over-estimate of the particle-hole channel
contributions. Therefore, a proper treatment should include the gap
effect in the particle-hole susceptibility.  In addition, the lowest
order treatment is no longer appropriate away from the weak coupling
regime.

Furthermore, it has now been established that as the pairing
interaction increases, pseudogap develops naturally
\cite{LeggettNature,Reviews2}. Experimental evidence for its existence
comes from high $T_{\text{c}}$ superconductors
\cite{ChenArcs,Reviews2,Timusk,arpesstanford_review} as well as atomic
Fermi gases
\cite{ThermoScience,heyan,Chen_MRRF,Jin6,JinStrinati_nphys}. Therefore,
a theory with proper treatment of the pseudogap effect is necessary in
order to arrive at results that can be \emph{quantitatively} compared
with experiment. For the same reason, the effect of the particle-hole
channel needs also to be studied within such a theory.

In this paper, we explore the particle-hole channel effect based on a
pairing fluctuation theory \cite{Chen1,Chen2}, originally developed
for treating the pseudogap phenomena of high $T_{\text{c}}$
superconductors.  This theory has been successfully applied to atomic
Fermi gases and has been generating results that are in good agreement
with experiment \cite{Reviews2,ThermoScience,Chen_MRRF}. Here we
include the entire particle-hole $T$-matrix, with gap effect included
in the fermion Green's functions. Instead of a simple average of the
particle-hole susceptibility $\chi_{\text{ph}}$ on the Fermi surface,
here we choose to average at two different levels -- one on the Fermi
surface, one over a narrow momentum shell around the Fermi level. We
find that $\chi_{\text{ph}}$ has very strong frequency and momentum as
well as temperature dependencies. It is sensitive to the gap
size. Therefore, self-consistently including the self-energy feedback
is important. For both levels of average, we find that while in the
BCS limit, the particle-hole channel effect may be approximated by a
downshift in the pairing strength so that the ratio
$2\Delta(0)/T_{\text{c}}$ is unaffected, the situation becomes more
complex as the interaction becomes stronger where the gap is no longer
very small. Significant difference exists for these two levels of
averaging. The particle-hole susceptibility is reduced by both a
smaller Fermi surface and a big (pseudo)gap in the crossover
regime. Deep in the BEC regime, the particle-hole channel
contributions drop to zero. Without including the incoherent part of
the self energy, we find that at unitarity, $T_{\text{c}}/E_{\text{F}}
\approx 0.217$, in reasonable agreement with experiment. 

The rest of this paper is arranged as follows. In Sec. II, we first
give a brief summary of the pairing fluctuation theory without the
particle-hole channel effect. Then we derive the theory with
particle-hole channel included, starting by studying the dynamic
structure of the particle-hole susceptibility.  Next, in Sec. III, we
present numerical results, showing the effect of the particle-hole
channel on the zero $T$ gap, transition temperature $T_{\text{c}}$ and
pseudogap at $T_{\text{c}}$, as well as the mean-field ratio
$2\Delta(0)/T_{\text{c}}^{\text{MF}}$. We also discuss and compare our
value of $T_{\text{c}}/E_{\text{F}}$ with experiment and those in the
literature. Finally, we conclude in Sec. IV. More detailed results on
the dynamic structure of the particle-hole susceptibility are
presented in the Appendix.

\section{Pairing fluctuation theory with the
  particle-hole channel effect included}
\label{sec:Theory}

\subsection{Summary of the pairing fluctuation theory without the
  particle-hole channel effect}
\label{subsec:Summary}

To make this paper self-contained and to introduce the assumptions as
well as the notations, we start by summarizing the pairing fluctuation
theory \cite{Chen1,Chen2} without the effect of the particle-hole
channel, as a foundation for adding the particle-hole channel. 


\begin{figure}[tb]
\centerline{\includegraphics[width=3.in]{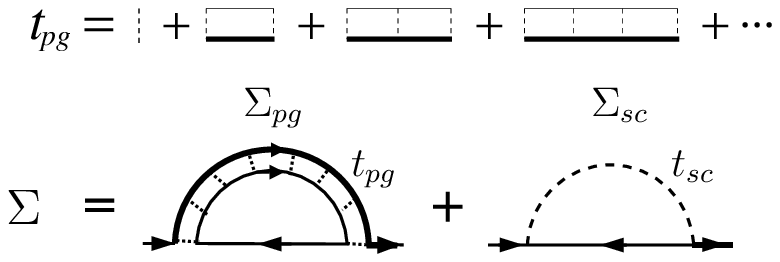}}
\caption{Feynman diagrams for the particle-particle channel $T$-matrix
  $t_{\text{pg}}$ and the self energy $\Sigma(K)$. The dotted lines represent
  the bare pairing interaction $U$. The dashed line, $t_{\text{sc}}$,
  represents the superfluid condensate.}
\label{fig:SelfE_Tmatrix}
\end{figure}


We consider a Fermi gas with a short range $s$-wave interaction
$U(\mathbf{k},\mathbf{k}')=U <0$, which exists only between opposite
spins. 
Our theory can be effectively represented by a $T$-matrix
approximation, shown diagrammatically in
Fig.~\ref{fig:SelfE_Tmatrix}. However, we emphasize that this is
\emph{not} a diagrammatic approach, since Fig.~\ref{fig:SelfE_Tmatrix}
is simply a representation of the equations derived from an equation
of motion approach \cite{Kadanoff,Patton1971,Chen1,ChenPhD}. The self
energy $\Sigma(K)$ comes from two contributions, associated with the
superfluid condensate and finite momentum pairs, respectively, given
by $\Sigma(K) = \Sigma_{\text{sc}}(K) + \Sigma_{\text{pg}}(K)\,$,
where
\begin{equation}
  \Sigma_{\text{sc}}(K) = \frac{\Delta_{\text{sc}}^2}{\i\omega_l +
    \ek}\,, \quad \Sigma_{\text{pg}}(K) = \sum_Q
  t_{\text{pg}}(Q)G_0(Q-K) \,,
\end{equation}
with $\Delta_{\text{sc}}$ being the superfluid order parameter.  We
use a four-vector notation, $K\equiv({\mathbf{k}}, \i\omega_l)$,
$Q\equiv({\mathbf{q}}, i\Omega_n)$, $\sum_K \equiv
T\sum_{l,\mathbf{k}}$, etc., and $\omega_l$ and $\Omega_n$ are odd and
even Matsubara frequencies for fermions and bosons, respectively. Here
$G_0(K)=(\i\,\omega_l - \ek)^{-1}$ and $G(K) = [G_0^{-1} -
\Sigma(K)]^{-1}$ are the bare and full Green's functions,
respectively, $\ek = \hbar^2k^2/2m - \mu$ is the free fermion
dispersion, measured with respect to the Fermi level. In what follows,
we will set $k_\text{B} = \hbar=1$. The pseudogap $T$-matrix
\begin{equation} 
t_{\text{pg}}(Q) = \frac{U}{1+U\chi(Q)} 
\label{eq:T_pg}
\end{equation}
can be regarded as the renormalized pairing interaction with pair
momentum $Q$, where
\begin{equation}
\chi(Q) = \sum_K G(K)G_0(Q-K)
\end{equation}
is the pair susceptibility. We emphasize that this asymmetric form of
$\chi(Q)$ is not an \textit{ad hoc} choice but rather a natural result
of the equation of motion method. The bare Green's function $G_0$
comes from the inversion of the operator $G_0^{-1}$ which appears on
the left hand side of the equations of motion. It also appears in the
particle-hole susceptibility $\chi_{\text{ph}}$, as will be shown
below. Albeit not a phi-derivable theory, the equation of motion
method ensures that this theory is more consistent with the
Hamiltonian than a phi-derivable theory.

The gap equation is given by the pairing instability condition, 
\begin{equation}
1+U\chi(0)=0, \qquad (T\le T_\text{c}),
\label{eq:Instability}
\end{equation}
referred to as the Thouless criterion, which can also be naturally
interpreted as the Bose condensation condition for the pairs, since
$1+U\chi(0) \propto \mu_{\text{pair}}$.  In fact, after analytical
continuation $\i\Omega_n \rightarrow \Omega+\i0^+$, one can Taylor
expand the (inverse) $T$-matrix as
\begin{equation}
t_{\text{pg}}(\Omega, \mathbf{q}) \approx  \frac{Z^{-1}}{\Omega -
  \Omega_\mathbf{q} + \mu_{\text{pair}} +\i\Gammaq}\;,
\label{eq:t_expansion}
\end{equation}
and thus extract the pair dispersion $\Omega_\mathbf{q} \approx
q^2/2M^*$ to the leading order, where $M^*$ is the effective pair
mass. Here $\Gammaq$ is the imaginary part of the pair dispersion and
can be neglected when pairs become (meta)stable
\cite{ChenPhD,Chen1,Chen2}. In the superfluid phase,
$t_{\text{pg}}(Q)$ diverges at $Q=0$ and a macroscopic occupation of
the $Q=0$ Cooper pairs, i.e., the condensate, appears. This
macroscopic occupation, has to be expressed as a singular term,
$t_{\text{sc}}(Q) = -(\Delta_{\text{sc}}^2/T)\delta(Q)$, (the dashed
line in Fig.~\ref{fig:SelfE_Tmatrix}), such that
$\Sigma_{\text{sc}}(K) = \sum_Q t_{\text{sc}}(Q) G_0(Q-K)$, written in
the same form as its pseudogap counterpart, $\Sigma_{\text{pg}}(K)$.

Now we split $\Sigma_{\text{pg}}(K)$ into coherent and incoherent parts:
\begin{eqnarray}
  \Sigma_{\text{pg}}(K) &=& \sum_Q \frac{t_{\text{pg}}(Q)} {\i\Omega_n
    -\i\omega_l-\xi_\mathbf{q-k}} \nonumber\\
  &=& -\sum_Q \frac{t_{\text{pg}}(Q)} {\i\omega_l+\xi_\mathbf{k}} +\delta\Sigma
  = \frac{\Delta_{\text{pg}}^2}{\i\omega_l+\ek} + \delta \Sigma 
\,,
\label{eq:Sigma_pg}
\end{eqnarray}
where we have defined the pseudogap $\Delta_{\text{pg}}$ via
\begin{equation}
\Delta_{\text{pg}}^2 \equiv -\sum_Q t_{\text{pg}}(Q) \approx Z^{-1} \sumq b(\Omegaq) \,,
\label{eq:PG}
\end{equation}
where $b(x)$ is the Bose distribution function.
Below $T_{\text{c}}$, the divergence of $t_{\text{pg}}(Q=0)$ makes it
a good mathematical approximation to neglect the incoherent term
$\delta\Sigma$ so that
%
\begin{equation}
\Sigma(K) \approx \frac{\Delta^2} {\i\omega_l+\ek} \,, \qquad
\text{with} \qquad  \Delta^2 =
\Delta_{\text{sc}}^2 +\Delta_{\text{pg}}^2 \,.
\label{eq:Sigma}
\end{equation}
%
Therefore, the Green's function $G(K)$, the quasiparticle dispersion
$\Ek = \sqrt{\ek^2 + \Delta^2}$, and the gap equation, as expanded
from Eq.~(\ref{eq:Instability}), follow the same BCS form,
\emph{except that the total gap $\Delta$ now contains both
  contributions from the order parameter $\Delta_{\text{sc}}$ and the
  pseudogap $\Delta_{\text{pg}}$.} 
%

For a contact potential, we get rid of the  interaction $U$ in
favor of the scattering length $a$ via ${m}/{4\pi a} = 1/U
+\sumk (1/{2\epsilon_\mathbf{k}})$, where $\epsilon_\mathbf{k} =
k^2/2m$. 
Then the gap equation can be written as
\begin{equation}
-\frac{m}{4\pi a} = \sumk \left[ \frac{1-2f(\Ek)}{2\Ek}
  -\frac{1}{2\epsilon_\mathbf{k}} \right]  \,,
\label{eq:gap}
\end{equation}
where $f(x)$ is the Fermi distribution function.  In addition, we have
the particle number constraint, $n = 2\sum_K G(K)$, i.e.,
\begin{equation}
n = 2\sumk \left[ \vk^2 + \frac{\ek}{\Ek}\,f(\Ek) \right] \,,
\label{eq:number}
\end{equation} 
where $\vk^2 = (1-\ek/\Ek)/2$ is the BCS coherence factor.

Equations (\ref{eq:gap}), (\ref{eq:number}), and (\ref{eq:PG}) form a
closed set. For given interaction $1/k_{\text{F}}a$, they can be used to solve
self consistently for $T_{\text{c}}$ as well as $\Delta$ and $\mu$ at $T_{\text{c}}$, or
for $\Delta$, $\Delta_{\text{sc}}$, $\Delta_{\text{pg}}$, and $\mu$ as a function of
$T$ below $T_{\text{c}}$. Here $k_{\text{F}}$ is the Fermi wave vector. More details about
the Taylor expansion of the inverse $T$ matrix can be found in
Refs.~\cite{ChenPhD,HePRB2007}.


\subsection{Dynamic structure of particle-hole susceptibility $\chi_{\text{ph}}(P)$}
\label{subsec:DynamicStructure}

\begin{figure}[tb]
%
\centerline{\includegraphics[]{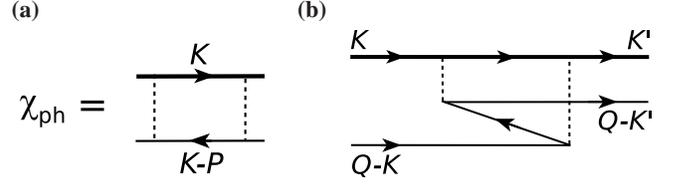}}
\caption{Feynman diagrams for the particle-hole susceptibility
  $\chi_{\text{ph}}$ in the presence of self-energy feedback effect. Panel
  (b) is identical to panel (a), with twisted external legs. The total
  particle-hole momentum $P$ in (a) is equal to $K+K'-Q$ in (b), with
  $Q$ being the particle-particle pair momentum.  }
\label{fig:1}
\end{figure}


Before we derive the theory with full particle-hole $T$-matrix
$t_\text{ph}$ included, we first study the dynamic structure of the
particle-hole susceptibility $\chi_{\text{ph}}(P)$. It is the single
rung of the particle-hole scattering ladder diagrams, as shown in
Fig.~\ref{fig:1}(a). Note that direct interaction exists only between
fermions of opposite spins. Therefore, the particle and hole must also
have opposite spins. The total particle-hole four-momentum is given by
$P \equiv (\i\nu_n, \mathbf{p})$. Since we are considering the effect
on pairing induced by the particle-hole channel, we can twist external
legs of the diagram, as shown in Fig.~\ref{fig:1}(b), so that the
particle-hole contribution can be added to the original pairing
interaction $U$ directly. It is obvious that the particle-hole
momentum $P$ in Fig.~\ref{fig:1}(a) are equal to $ K+K'-Q$ in
Fig.~\ref{fig:1}(b), where $Q$ is the pair momentum of the
particle-particle channel. Therefore, we have
%
\begin{eqnarray}
\lefteqn{\chi_{\text{ph}}(P) 
 = \sum_K G(K) G_0(K-P)} \nonumber\\
&&{}= \sumk \left[ \frac{f(\Ek)-f(\xi_\mathbf{k-p})}
  {\Ek-\xi_\mathbf{k-p}-i\nu_n} \uk^2
  -\frac{1-f(\Ek)-f(\xi_\mathbf{k-p})}
  {\Ek+\xi_\mathbf{k-p}+i\nu_n}\vk^2 \right]\,. \nonumber\\
\label{eq:chi_ph}
\end{eqnarray}
Note that again we have a mixing of dressed and undressed Green's
function in $\chi_{\text{ph}}(P)$, like in the expression of
$\chi(Q)$. As mentioned earlier, this mixing has exactly the same
origin in both cases \cite{NotesonG0Gph}. For convenience, here we
dress the particle propagator with self energy and leave the hole
propagator undressed. This is based on the fact that the hole
propagator is undressed in $\chi(Q)$ in
Sec.~\ref{subsec:Summary}. (One can equivalently dressed the hole
while leaving the particle undressed).

A few remarks are in order. Firstly, in the center-of-mass (COM)
reference frame of a particle-particle pair, the momenta in
Fig.~\ref{fig:1}(b) will be relabeled as $\pm K+Q/2$ and $\pm K'+Q/2$,
so that $P = (K+Q/2) - (-K'+Q/2) = K+K'$, independent of the total
pair momentum $Q$. Here $\pm K$ and $\pm K'$ are the four momenta of
the incoming and outgoing fermions in the COM reference
frame. Therefore, the induced interactions still conform to the
Galileo transformation.  Secondly, as in the Nozi\'eres and
Schmitt-Rink (NSR) theory \cite{NSR}, one needs a fictitious separable
potential $U_\mathbf{k,k'} = U \varphi_\mathbf{k}\varphi_\mathbf{k'}$
in order to have a simple result in the form of Eq.~(\ref{eq:T_pg})
for the summation of the particle-particle ladder diagrams
\cite{Janko,Chen1,Chen2}. The contact potential considered for atomic Fermi
gases automatically satisfies this requirement. However, since the
particle-hole pair susceptibility $\chi_{\text{ph}}(P)$ only depends
on the sum $P=K+K'$, it is obvious that when the particle-hole
contribution is included, the total effective interaction
$U_{\text{eff}}(\mathbf{k,k'})$ will no longer be separable.
Approximate treatment is needed, as will be shown later.

To proceed, we separate the retarded $\chi_{\text{ph}}^R$ into real and
imaginary parts, $\chi_{\text{ph}}^R(\nu, \mathbf{p}) =
\chi^\prime_{\text{ph}}(\nu,\mathbf{p})+
\i\chi^{\prime\prime}_{\text{ph}}(\nu,\mathbf{p})$, after analytical
continuation, $\i\nu_n \rightarrow \nu + \i0^+$. We have
\begin{subequations}
\begin{eqnarray}
\lefteqn{\chi^\prime_{\text{ph}}(\nu,\mathbf{p})}\nonumber\\
&&\hspace*{1ex} {} = \sumk \left[ \frac{f(\Ek)-f(\xi_\mathbf{k-p})}
  {\Ek-\xi_\mathbf{k-p}-\nu} \uk^2
  -\frac{1-f(\Ek)-f(\xi_\mathbf{k-p})}
  {\Ek+\xi_\mathbf{k-p}+\nu}\vk^2 \right]\,,\nonumber\\
\\
\lefteqn{\chi^{\prime\prime}_{\text{ph}}(\nu,\mathbf{p})}\nonumber\\
&&\hspace*{1ex}{} = \pi\sumk\left\{ \left[ f(\Ek)-f(\Ek-\nu)\right]
  \uk^2\delta(\Ek-\xi_\mathbf{k-p}-\nu) \right.\nonumber\\
&&\hspace*{1.5cm}{}\left.  + \left[f(\Ek+\nu)-f(\Ek)\right] \vk^2
  \delta(\Ek+\xi_\mathbf{k-p}+\nu) \right\}\,.\nonumber\\
\label{eq:chiph_im}
\end{eqnarray}
\end{subequations}
From Eq.~(\ref{eq:chiph_im}), we can immediately conclude
$\chi_{\text{ph}}^{\prime\prime}(0, \mathbf{p}) = 0$. In addition,
$\chi_{\text{ph}}^{\prime\prime}(\nu, 0) = 0$ if
$-\min(\Ek+\ek)=-(\sqrt{\mu^2+\Delta^2}-\mu) < \nu < 0$ or $\nu >
\max(\Ek-\ek) = \sqrt{\mu^2+\Delta^2}+\mu$. At low $T$, we also have
$\chi_{\text{ph}}^{\prime\prime}(\nu, 0)$ exponentially small for $|\nu|<
\Delta$ if $\mu>0$ or for $|\nu|<\sqrt{\mu^2+\Delta^2}$ otherwise. In
all cases, $\chi_{\text{ph}}^{R}(\nu, \mathbf{p}) =\chi_{\text{ph}}^{R}(\nu, p)$ is
isotropic in $\mathbf{p}$. In the BCS limit, $\Delta \rightarrow 0$, $\Ek
\rightarrow |\ek|$, so that
\begin{equation}
\chi_{\text{ph}}^{\prime}(0, p\rightarrow 0) \approx \sumk
f^\prime(\ek) = -\frac{mk_\mu}{2\pi^2} \,,
\end{equation}
where $k_\mu = \sqrt{2m\mu}$ is the momentum on the Fermi surface.

\begin{figure*}[tb]
\centerline{\includegraphics[width=5.5in]{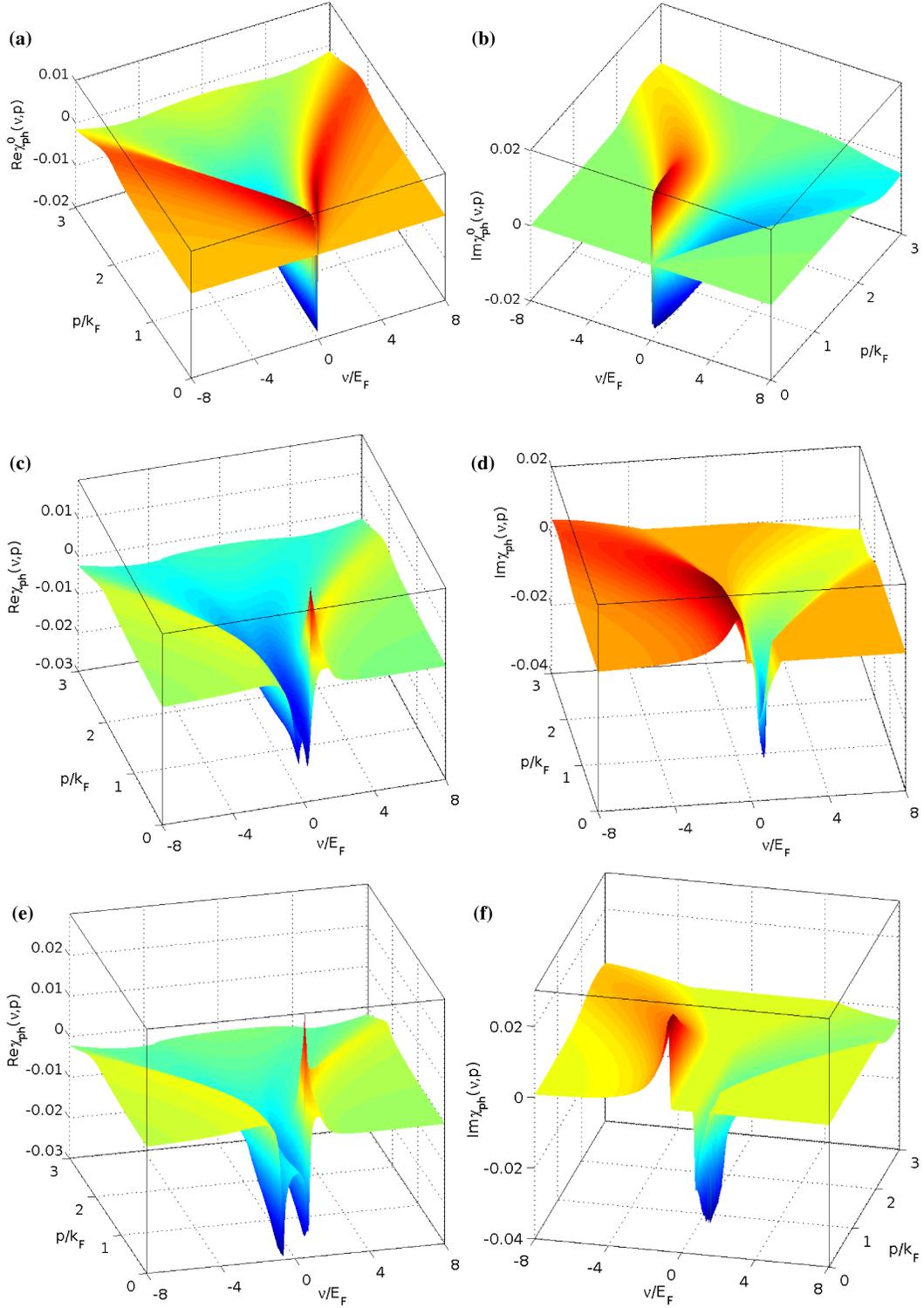}}
\caption{(Color online) 3D plots of the real (a,c,e) and imaginary
  (b,d,f) parts of the particle-hole pair susceptibility $\chi_{\text{ph}}$
  with (c-f) and without (a,b) the self-energy feedback, calculated at
  $T_{\text{c}}$ (a-d) and $0.1T_{\text{c}}$ (e,f) in the unitary limit. Here $T_{\text{c}}$ is
  calculated using the pairing fluctuation theory without the
  particle-hole channel contributions. While the undressed
  $\chi_{\text{ph}}^{0R}(\nu,p)$ has a simple symmetry under $\nu\rightarrow
  -\nu$, the dynamic structure of $\chi_{\text{ph}}^{R}(\nu,p)$ is much more
  complex. In both cases, the real and imaginary parts have very
  strong dependencies on the total frequency $\nu$ and total
  momentum $p$. $\chi_{\text{ph}}^{R}(\nu,p)$ shows strong gap effects both
  at $T_{\text{c}}$ and low $T$. In units of $E_{\text{F}}$, the parameters are: $T_{\text{c}} =
  0.256$, $\mu(T_{\text{c}}) = 0.62$, $\Delta(T_{\text{c}})=0.64$, $\mu(0.1T_{\text{c}}) = 0.59$,
  and $\Delta(0.1T_{\text{c}}) = 0.69$.}
\label{fig:chi_ph}
\end{figure*}

For comparison, we also study the dynamic structure of the undressed
particle-hole pair susceptibility, 
\begin{equation}
\chi_{\text{ph}}^0(P) = \sum_K G_0(K)G_0(K-P) =\sumk
\frac{f(\ek)-f(\xi_\mathbf{k-p})} { \ek - \xi_\mathbf{k-p} -\i\nu_n} \,.
\end{equation}
This is the particle-hole pair susceptibility studied by GMB
\cite{GMB} and others
\cite{Heiselberg2000,KimTorma2009,TormaPethick2009,Yin2009} in the
literature. The imaginary part is given by
\begin{equation}
\chi_{\text{ph}}^{0\prime\prime}(\nu,\mathbf{p}) = \pi \sumk \left[ f(\ek) - f(\ek
  -\nu)\right] \delta (\ek -\xi_\mathbf{k-p} -\nu)\,.
\label{eq:Imchi0_ph}
\end{equation}
Same as in the $\chi_{\text{ph}}(P)$ case, we have
$\chi_{\text{ph}}^{0\prime\prime}(0,\mathbf{p}) = 0$. For finite $\nu \neq
0$,  
$\chi_{\text{ph}}^{0\prime\prime}(\nu,\mathbf{p}) \rightarrow 0$
exponentially as $p \rightarrow 0$. 
The real part satisfies
\begin{equation}
\chi_{\text{ph}}^{0\prime}(\nu \neq 0, 0) = 0
\label{eq:Rechi0_p=0}
\end{equation}
and 
\begin{equation}
\chi_{\text{ph}}^{0\prime}(0, p\rightarrow 0) = \sumk f'(\ek) \approx
-\frac{mk_\mu}{2\pi^2}
\label{eq:Rechi0_nu=0}
\end{equation}
at low $T$. More generally, for $\nu=0$ and finite $p$, we have
\begin{equation}
\chi_{\text{ph}}^{0\prime}(0, p) = \int_0^\infty \frac{k\d k}{2\pi^2}
\frac{m}{p} f(\ek) \ln \left| \frac{2k-p} {2k+p}\right| \,.  
\label{eq:Rechi0_nu=0_2}
\end{equation}
In the weak coupling limit, $\chi_{\text{ph}}^{\prime}(0, p\rightarrow
0) = \chi_{\text{ph}}^{0\prime}(0, p\rightarrow 0) $, since
$\chi_{\text{ph}}$ reduces to $\chi_{\text{ph}}^{0}$ when the gap
$\Delta$ vanishes.

Let's take a look at the case of \emph{small} but finite $p$. Equation
(\ref{eq:Imchi0_ph}) can be rewritten as
\begin{eqnarray}
\chi_{\text{ph}}^{0\prime\prime}(\nu,\mathbf{p}) &=& \pi \sumk \left[ f(\ek) - f(\ek
  -\nu)\right] \delta \left(\frac{\mathbf{k}\cdot
    \mathbf{p}}{m}-\frac{p^2}{2m} -\nu\right)\,\nonumber\\
& \approx &  \pi\nu \sumk  f'(\ek) \delta \left(\frac{\mathbf{k}\cdot
    \mathbf{p}}{m}-\frac{p^2}{2m} -\nu\right) \propto \nu \,.
\end{eqnarray}
%
The delta function can be satisfied only at $k \gtrsim |\nu|
m/p$. When $|\nu|m/p > k_\mu$, $\ek > 0$ for all $k$ 
so that the magnitude of
$\chi_{\text{ph}}^{0\prime\prime}(\nu,\mathbf{p})$ will also turn
around and start to decrease exponentially. The turning points $\nu
= \pm pk_\mu/m$ show up as two peaks in 
$\chi_{\text{ph}}^{0\prime}(\nu,\mathbf{p})$.

It is easy to show that the hermitian conjugate
$\chi_{\text{ph}}^{0R*}(\nu, \mathbf{p}) = \chi_{\text{ph}}^{0R}(-\nu,
\mathbf{p})$.
%
%
Similar relations do not hold for $\chi_{\text{ph}}$, however, due to
the mixing of $G_0$ and $G$ in the expression of
$\chi_{\text{ph}}(P)$.

Shown in Figs.~\ref{fig:chi_ph}(a) and \ref{fig:chi_ph}(b) are three
dimensional (3D) plots of the real and imaginary parts of
$\chi_{\text{ph}}^{0R}(\nu, \mathbf{p})$. In Figs.~\ref{fig:chi_ph}(c)
through \ref{fig:chi_ph}(f) we present our calculated results of
$\chi_{\text{ph}}^{R}(\nu, \mathbf{p})$ in the presence of self energy
feedback. They are calculated at $T_{\text{c}}$ (a-d) and
$0.1T_{\text{c}}$ (e,f) in the unitary limit,
$1/k_{\text{F}}a=0$. Here $T_{\text{c}}/E_{\text{F}}\approx 0.256$ is
the one calculated in the pairing fluctuation theory without including
the particle-hole channel contribution. The even and odd symmetries of
$\chi_{\text{ph}}^{0\prime}(\nu, \mathbf{p})$ and
$\chi_{\text{ph}}^{0\prime\prime}(\nu, \mathbf{p})$ with respect to
$\nu \rightarrow -\nu$ are evident. And indeed these symmetries are
not present for $\chi_{\text{ph}}^{R}(\nu, \mathbf{p})$. The
interesting structure at low frequency and low momentum clearly
derives from the pseudogap already present at $T_{\text{c}}$ in the
fully dressed Green's function. In other words, by neglecting the
feedback effect, the bare $\chi_{\text{ph}}^0(P)$ misses this
important dynamic structure. The plots of $\chi_{\text{ph}}^{0R}$ at
$0.1T_{\text{c}}$ (not shown) is very similar to its $T=T_{\text{c}}$
counterpart shown in Figs.~\ref{fig:chi_ph}(a) and
\ref{fig:chi_ph}(b), except that the peaks become sharper. Comparing
Figs.~\ref{fig:chi_ph}(e) and \ref{fig:chi_ph}(f) with
\ref{fig:chi_ph}(c) and \ref{fig:chi_ph}(d), the gap induced
structures become much more pronounced. For example, for $p=0$, the
range of $\nu$ in which $\chi_{\text{ph}}^{\prime\prime}(\nu, 0)=0$
becomes much wider at low $T$.

More quantitatively readable two-dimensional plots are presented in
the Appendix. There we study in detail the effect of temperature and
interaction strength on the particle-hole susceptibility, and how it
behaves as a function of frequency $\nu$ for fixed momentum $p$ or as
a function of total momentum $p$ for fixed frequency $\nu$.

\begin{figure}[tb]
\centerline{\includegraphics[width=3.2in,clip]{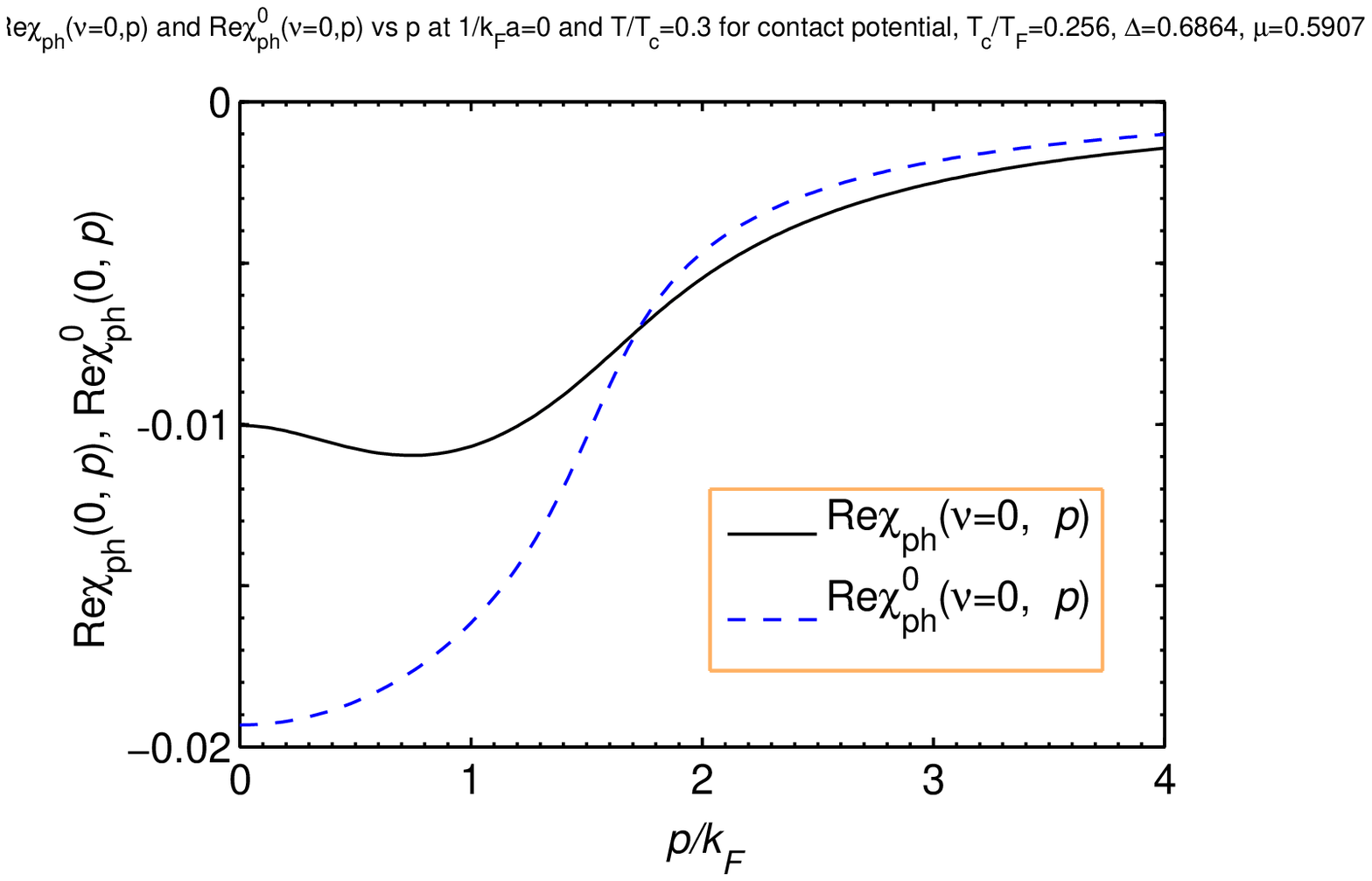}}
\caption{(Color online) Strong momentum dependence of the real part of the
  particle-hole susceptibility at zero frequency $\nu=0$ in the
  unitarity limit, with (black curve) and without (blue dashed curve)
  self-energy feedback, calculated at $T=0.3T_{\text{c}}$, where $T_{\text{c}} =
  0.256E_{\text{F}}$. While the undressed $\chi_{\text{ph}}^{0\prime}(0,p) =
  \Re\chi_{\text{ph}}^{0}(0,p)$ shows a simple monotonic behavior, the
  dressed susceptibility $\chi_{\text{ph}}^{\prime}(0,p) = \Re\chi_{\text{ph}}(0,p)$
  has a nonmonotonic $p$ dependence, and a substantially reduced value
  at $p=0$. This reduction derives from the gap effect in the Green's
  function $G(K)$. Namely, $\chi_{\text{ph}}^{\prime}(0,p)$ seriously
  over-estimated particle-hole fluctuations.}
\label{fig:chi_ph_nu0-p}
\end{figure}

In Fig.~\ref{fig:chi_ph_nu0-p}, we plot systematically the zero
frequency value of the real part of the particle-hole pair susceptibility
as a function of total momentum $p$, with and without the feedback
effect. The curves are computed at a relatively low $T=0.3T_{\text{c}}$ at
unitarity. Due to the large excitation gap $\Delta=0.69E_{\text{F}}$, at $p=0$,
the value $\chi_{\text{ph}}^{\prime}(0,0)$ with the feedback is strongly
suppressed from its undressed counterpart,
$\chi_{\text{ph}}^{0\prime}(0,0)$. In other words, \emph{the neglect of the
  self-energy feedback in $\chi_{\text{ph}}^{0\prime}(0,0)$ leads to serious
  over-estimate of the particle-hole channel contributions}. At the
same time, $\chi_{\text{ph}}^{\prime}(0,p)$ exhibits a more complex,
nonmonotonic dependence on $p$ than $\chi_{\text{ph}}^{0\prime}(0,p)$. In
both cases, the momentum dependence is strong.

From Figs.~\ref{fig:chi_ph} and \ref{fig:chi_ph_nu0-p}, as well as the
Appendix, one can readily see that the particle-hole susceptibility
$\chi_{\text{ph}}^R(\nu,p)$ has very strong dependencies on both frequency
and momentum, as well as the temperature and interaction strength.

\subsection{Induced interaction -- beyond the lowest order}
\label{subsec:InducedInteraction}

Figure \ref{fig:chi_ph}(b), in the absence of the self-energy feedback
effect, is in fact the lowest order induced interaction, considered in
GMB and most others in the literature:
\begin{equation}
U^0_{\text{ind}}(P) = -U^2 \chi^0_{\text{ph}}(P) .
\end{equation}
Diagrams of the same order but between fermions of the same spin
vanish. 

\begin{figure}[tb]
\centerline{\includegraphics[]{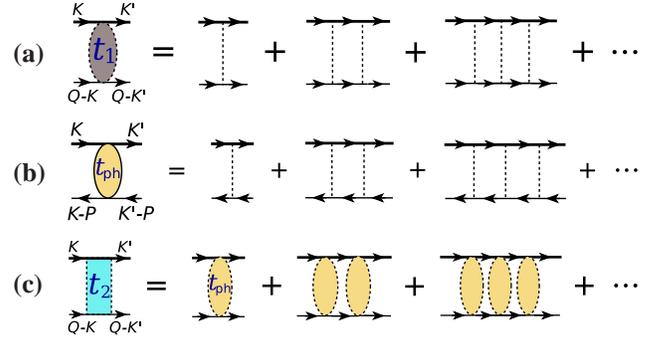}}
\caption{(Color online) Feynman diagrams showing the particle-hole
  effect on fermion pairing, in the presence of self-energy
  feedback. (a) Particle-particle $T$ matrix $t_1(Q)$, with external
  four momenta labeled. (b) Particle-hole $T$ matrix $t_{\text{ph}}(P)$, with
  $P=K+K'-Q$ being the total particle-hole 4-momentum. (c) An
  effective, composite particle-particle $T$-matrix, $t_2(Q)$, with
  the contribution from the particle-hole channel included. Here
  different shadings represent different $T$ matrices.}
\label{fig:t_2}
\end{figure}

Let us first re-plot the particle-particle scattering $T$-matrix in
Fig.~\ref{fig:t_2}(a); this is just $t_{\text{pg}}$ in
Fig.~\ref{fig:SelfE_Tmatrix} but with external legs, which
we here refer to as $t_1(Q)$. We have 
\begin{equation}
t_1(Q) = \frac{1}{U^{-1}+\chi(Q)}\,.
\label{eq:t_1}
\end{equation}
Now we consider the contribution of an infinite particle-hole ladder
series, as shown in Fig.~\ref{fig:t_2}(b). Such contribution should be
added to the bare interaction $U$. The summation of the series of
particle-hole ladder diagrams gives rise to the $T$-matrix in the
particle hole channel,
\begin{equation}
t_{\text{ph}}(P) =  \frac{U}{1+U\chi_{\text{ph}}(P)} = \frac{1}{U^{-1} + \chi_{\text{ph}}(P)} \,.
\label{eq:t_ph}
\end{equation}
At $Q=0$, this gives the overall effective pairing interaction,
\begin{equation}
U_{\text{eff}}(K,K') = t_{\text{ph}}(K+K') = U + U_{\text{ind}} = \frac{U}{1+U\chi_{\text{ph}}(K+K')} \;,
\end{equation}
where $K$ and $K'$ are the incoming and outgoing 4-momenta of the
scattering particles in the COM reference frame. The
induced interaction is thus given by
\begin{equation}
U_{\text{ind}}(P) = t_{\text{ph}}(P) - U = - \frac{U^2\chi_{\text{ph}}(P)}
{1+U\chi_{\text{ph}}(P)} \,,
\end{equation}
with $P=K+K'$. Upon Taylor expanding the denominator by treating
$U\chi_{\text{ph}}$ as a small parameter, one readily notice that the leading
term is just the numerator. This would be the counterpart lowest
induced interaction in our theory, except that we always consider the
self energy feedback effect.

It is evident that the $T$ matrices in the particle-particle channel
and the particle-hole channel share the same lowest order term,
$U$. Both $T$ matrices can be regarded as a renormalized interaction,
but in different channels. What we need is to replace the bare $U$ in
one of the two $T$ matrices with the other $T$ matrix. The results are
identical, which we call $t_2$. Shown in Fig.~\ref{fig:t_2}(c) is the
regular particle-particle channel $T$ matrix $t_1(Q)$ with $U$
replaced by the particle-hole channel $T$ matrix $t_{\text{ph}}(P)$ (with
twisted external legs), where $P=K+K'-Q$. In other words, we replace
$U^{-1}$ with $t_{\text{ph}}^{-1}(P) = U^{-1} + \chi_{\text{ph}}(P)$ in
Eq.~(\ref{eq:t_1}), and formally obtain
\begin{equation}
t_2(Q) = \frac{1}{U^{-1} + \chi_{\text{ph}}(K+K'-Q) + \chi(Q)} \,.
\label{eq:t_2}
\end{equation}
%
Unfortunately, since $U_{\text{eff}}(K,K')$ is \emph{not} a separable
potential, one \emph{cannot} obtain a simple summation in the form of
Eq.~(\ref{eq:t_2}). This can also be seen from the extra dependence on
$K$ and $K'$ on the right hand side of the equation. Certain averaging
process has to be done to arrive at such a simple summation, as will
be shown below.

\subsection{Gap equation from self-consistency condition in mean-field
treatment}
\label{subsec:mean-field}

The dependence of $U_{\text{eff}}(P)$ on external momenta via $P=K+K'-Q$
presents a complication in the gap equation. This can be seen through
the self consistency condition in the mean field treatment, even
though we do not use mean field treatment in our calculations. Writing
the interaction $V_{K,K'} = U_{\text{eff}}(K+K')$ for $Q=0$, i.e., zero total
pair 4-momentum, we have
\begin{eqnarray}
\Delta_K &=& \sum_{K'} V_{K,K'} \langle c^{}_{K'}c_{-K'}^{}\rangle
\nonumber\\
&=& -\sum_{K'} \frac{U}{1+U\chi_{\text{ph}}(K+K')}
\frac{\Delta_{K'}}{(\i\omega_{l'})^2 -E_{K'}^2} \,,
\end{eqnarray}
where we have used the mean-field result $\langle
c^{}_{K'}c_{-K'}^{}\rangle = G(K')G_0(-K')\Delta_{K'}$. Equivalently,
this can be written as
\begin{equation}
\Delta_{\mathbf{k},\i\omega_l} = -\sum_{K'} \frac{U}
{1+U\chi_{\text{ph}}(\i\omega_l+\i\omega_{l'}, \mathbf{k+k'})} \;
\frac{\Delta_{\mathbf{k'},\i\omega'}} {(\i\omega')^2 -E_{K'}^2} \,.
\end{equation}
Note that, due to the dynamic character of $\chi_{\text{ph}}(K+K')$, both the
gap $\Delta_K$ and the quasiparticle dispersion $E_K$ acquire a
dynamical frequency dependence. The gap also develops a momentum
dependence, which is originally absent for a contact potential.

We can express $U_{\text{eff}}(P)$ in terms of its retarded analytical
continuation, as follows:
\begin{equation}
U_{\text{eff}}(P) = U + \int^\infty_{-\infty} \frac{\d \nu}{2\pi} \;
\frac{-2\,\mbox{Im}\, U_{\text{eff}}^R(\nu,\mathbf{p})} {\i\nu_n -\nu} \,,
\end{equation}
where the second term is just the induced interaction, 
\begin{equation}
\mbox{Im}\, U_{\text{eff}}^R(\nu,\mathbf{p}) =
\frac{\chi_{\text{ph}}^{\prime\prime}(\nu,\mathbf{p})}
{(U^{-1}+\chi_{\text{ph}}^{\prime})^2 + (\chi_{\text{ph}}^{\prime\prime})^2} \,.
\end{equation}
 Then we have
\begin{equation}
\Delta_{\mathbf{k},\i\omega_l} = U
\sum_{K'}\frac{\Delta_{\mathbf{k'},\i\omega_{l'}}} {(\i\omega_{l'})^2 -E_{K'}^2}
 -\int^\infty_{-\infty} \frac{\d \nu}{\pi} \; \frac{\mbox{Im}\,
  U_{\text{eff}}^R(\nu,\mathbf{k+k'})} {\i\omega_l+\i\omega_{l'} -\nu}\,.  
\end{equation}
The particle-hole channel effect is contained in the 2nd term, without
which this would be just the gap equation without the particle-hole
channel, and admit a constant gap solution.
Without further approximation, the complex dynamic structure of
$\chi_{\text{ph}}(P)$ will inevitably render it very difficult to
solve the gap equation.

\subsection{Pairing instability condition in the presence of particle-hole
channel effect}
\label{subsec:PairingInstability}


In order to obtain a simple form as Eq.~(\ref{eq:t_2}), we have to
average out the dependence of $U_{\text{eff}}(K,K')$ on $K$ and $K'$.
Indeed, an average of $\chi_{\text{ph}}(\nu,p)$ has been performed in
the literature on (and only on) the Fermi surface \cite{GMB}. For the
frequency part, here we follow the literature and take $\i\nu_n =
\i\omega_l + \i\omega_{l'} =0$.  From Fig.~\ref{fig:chi_ph}, one can
see that this is where the imaginary part
$\chi_{\text{ph}}^{\prime\prime}(\nu,p) = 0$ for all $p$ and thus the
effective interaction $U_{\text{eff}}(K,K')$ is purely real. For the
momentum part, we choose on-shell, elastic scattering, i.e., $k = k'$,
and then average over scattering angles:
\begin{equation}
p = |\mathbf{k}+\mathbf{k'}| = k \sqrt{2\,(1+\cos \theta)} \;,
\end{equation}
where $\theta$ is the angle between $\mathbf{k}$ and $\mathbf{k'}$. It
is the off-shell scattering processes which lead to imaginary part and
nontrivial frequency dependence in $\chi_{\text{ph}}^{R}(\nu,p)$ and the
order parameter.
Further setting $k=k_\mu$ and averaging only on the Fermi surface is
the averaging process used in all papers we can find about induced
interactions in the literature. We refer to this as \emph{level 1}
averaging. In this paper, we also perform a \emph{level 2} average,
over a range of $k$ such that the quasiparticle energy $\Ek \in
[\min(\Ek), \min(\Ek)+\Delta]$. Here $\min(\Ek) = \Delta$ if $\mu>0$,
or $\min(\Ek) = \sqrt{\mu^2+\Delta^2}$ if $\mu<0$. The basic idea is
that according to the density of states of a typical $s$-wave
superconductor, the states within the energy range $\Ek \in [\Delta,
2\Delta]$ are most strongly modified by pairing. It should be pointed
out that in the BEC regime, this range can become very large.

Upon averaging of either level 1 or level 2, we drop out the
complicated dynamical structure of $\chi_{\text{ph}}(\nu,p)$ and replace it
by a constant $\langle \chi_{\text{ph}}\rangle$. For the purpose of
comparison, we shall also perform the averaging on the undressed
particle-hole susceptibility $\chi_{\text{ph}}^0(\nu,p)$ but will mostly show
the result at level 1.

\begin{figure}[tb]
\centerline{\includegraphics[width=3.2in,clip]{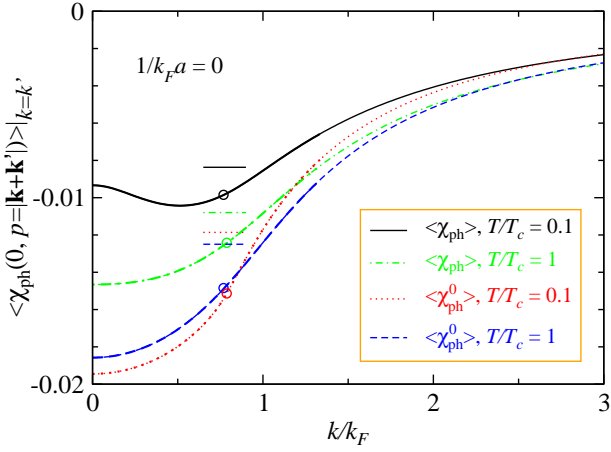}}
\caption{(Color online) Angular average of the on-shell particle-hole
  susceptibility, $\langle
  \chi_{\text{ph}}(0,p=|\mathbf{k}+\mathbf{k}'|)\rangle $ at $\nu=0$ as a
  function of momentum $k/k_{\text{F}}$, under the condition $k=k'$, calculated
  at unitarity for different temperatures $T=0.1T_{\text{c}}$ (black solid
  curve) and $T=T_{\text{c}}$ (green dot-dashed curve), in units of
  $k_{\text{F}}^3/E_{\text{F}}$.  Also plotted is its undressed counterpart, $\langle
  \chi_{\text{ph}}^0(0,p=|\mathbf{k}+\mathbf{k}'|)\rangle $, which shows a
  serious over-estimate due to the neglect of the self-energy
  feedback. Here $T_{\text{c}}=0.256E_{\text{F}}$ and associated gap and $\mu$ values
  are calculated without the particle-hole channel effect. The open
  circles on each curve denote level 1 average, i.e., $k=k_\mu$. The
  vertical axis readings of the horizontal short bars indicate the
  corresponding values of level 2 average. The thick section of each
  curve indicates the range of $k$ used for level 2
  averaging. Clearly, there are strong temperature and $k$
  dependencies in both $\langle \chi_{\text{ph}}(0,p)\rangle $ and $\langle
  \chi_{\text{ph}}^0(0,p)\rangle $. The (absolute) values of Level 2 average
  are substantially smaller than their level 1 counterpart. }
\label{fig:chi_ph_avg-k}
\end{figure}

Shown in Fig.~\ref{fig:chi_ph_avg-k} are the angular averages of the
particle-hole susceptibility at $\nu=0$ as a function of momentum $k$
under the above on-shell condition, $k=k'$. Here we only show the
unitary case at two different temperatures, $T=T_{\text{c}}$ and low $T=0.1T_{\text{c}}
\ll T_{\text{c}}$. For the purpose of comparison, we plot the result for both
the dressed and undressed particle-hole susceptibility. The curves
show strong momentum dependencies. For $\langle
\chi_{\text{ph}}^0(0,p)\rangle $, it is monotonically increasing, whereas for
$\langle \chi_{\text{ph}}(0,p)\rangle $, it exhibits nonmonotonic $k$
dependence at low $T$. Both dressed and undressed particle-hole
susceptibilities have a temperature dependence, and this dependence is
much stronger for the former.  This can be attributed mainly to the
temperature dependence of $\Delta(T)$ in $\langle
\chi_{\text{ph}}(0,p)\rangle $, while $\langle \chi_{\text{ph}}^0(0,p)\rangle $
depends on $T$ only via $\mu(T)$. 

The open circles on each curve represent the level 1 average, i.e., the
values at $k=k_\mu$. At the same time, the vertical axis readings of
the short horizontal bars correspond to the level 2 average, while the
thick segments of each curve represents the range of $k$ used for
level 2 averaging. Figure \ref{fig:chi_ph_avg-k} shows that the
(absolute) values of the level 2 average are significantly smaller
than their level 1 counterpart. The level 1 average $\langle
\chi_{\text{ph}}^0(0,p)\rangle$ is essentially temperature independent
(see the red and blue circles).
In addition, it is evident that \emph{the neglect of self-energy
  feedback has caused $\langle \chi_{\text{ph}}^0(0,p)\rangle $ to seriously
  over-estimate the contribution of particle-hole channel.}

Similar plot for $1/k_{\text{F}}a = 0.5$ (Fig.~\ref{fig:chi_ph_avg-k_BEC} in
the Appendix) exhibits a much stronger $T$ dependence. In that case,
$\mu$ is very close to 0 albeit still positive. As a consequence, the
particle-hole susceptibility is much smaller than that shown in
Fig.~\ref{fig:chi_ph_avg-k}.

Now with this frequency and momentum independent $\chi_{\text{ph}}(\nu,p)
\approx \langle \chi_{\text{ph}}\rangle $, we can easily carry out the simple
geometric summation for $t_2$:
\begin{equation}
t_2(Q) = \frac{1}{U^{-1} + \langle \chi_{\text{ph}}\rangle + \chi(Q) } \,.
\label{eq:t_2_avg}
\end{equation}
Therefore, the Thouless criterion for pairing instability leads to the
gap equation:
\begin{equation}
U^{-1} + \langle \chi_{\text{ph}}\rangle + \chi(0) = 0,
\label{eq:Thouless}
\end{equation}
namely, 
\begin{equation}
-\left(\frac{m}{4\pi a}+\langle \chi_{\text{ph}}\rangle\right) = \sumk \left[
  \frac{1-2f(\Ek)}{2\Ek} 
  -\frac{1}{2\epsilon_\mathbf{k}} \right]  \,.
\label{eq:gap_ph}
\end{equation}
As will be shown later, $\langle \chi_{\text{ph}}\rangle$ is always
negative. Therefore, the particle-hole channel effectively reduces
the strength of the pairing interaction. 

In the weak coupling limit ($1/k_{\text{F}}a = -\infty$), $\Delta \rightarrow
0$, $T\lesssim T_{\text{c}} \ll T_{\text{F}}$, then $\langle \chi_{\text{ph}}\rangle$ and
$\langle \chi_{\text{ph}}^0\rangle$ become equal, for either level of
averaging. We have
\begin{eqnarray}
\langle\chi_{\text{ph}}\rangle &=& \int_{-1}^1 \d x \int_0^\infty \frac{k\d k}{4\pi^2}
\frac{m}{p} f(\ek) \ln \left| \frac{2k-p} {2k+p}\right|_{p =k_{\text{F}}
  \sqrt{2\,(1+x)}} \nonumber\\
&\approx& N(0)\int_{-1}^1 \d x \int_0^1 \frac{\tilde{k}\d
  \tilde{k}}{2\tilde{p}} \ln
\left|\frac{2\tilde{k}-\tilde{p}}{2\tilde{k}+\tilde{p}}\right|
\nonumber\\
&=& -\frac{1+2\ln 2}{3} N(0) = 0.02015 \frac{k_{\text{F}}^3}{E_{\text{F}}}\,,  
\label{eq:Rechi0_avg}
\end{eqnarray}
where $\tilde{k}= k/k_{\text{F}}$, $\tilde{p} = p/k_{\text{F}} = \sqrt{2(1+x)}$, $x=\cos
\theta$, and $N(0) = mk_{\text{F}}/2\pi^2\hbar^2$ is the density of state at
the Fermi level. Here we have approximated the Fermi function with its
$T=0$ counterpart, with a step function jump at the Fermi level.

In the weak interaction limit, the BCS result for $T_{\text{c}}$ is
$T^{BCS}_{c}/E_{\text{F}} = (8/\pi)\,\e^{\gamma-2} \e^{1/N(0)U}$, where $\gamma
\approx 0.5772157$ is the Euler's constant.  Equation
(\ref{eq:Thouless}) implies a replacement of $1/U$ by
$1/U+\langle\chi_{\text{ph}}\rangle$. In this way, the new
transition temperature $T_{\text{c}}$ is given by
\begin{equation}
\frac{T_{\text{c}}\;\;\;\;}{ T_{\text{c}}^{BCS}} = \e^{\langle\chi_{\text{ph}}\rangle/N(0)} 
= (4\e)^{-1/3} \approx 0.45 \,,
\label{eq:gammaTc}
\end{equation}
and the same relation holds for zero $T$ gap,
\begin{equation}
\frac{\Delta\;\;\;\;\;}{\Delta^{BCS}} = (4\e)^{-1/3} \,.
\label{eq:gammaGap}
\end{equation}
This result is in \emph{quantitative} agreement (to the leading order)
with that of GMB \cite{GMB} and others \cite{Heiselberg2000} in the
literature. Note that in our work, as well as in that of Yin and
coworkers \cite{Yin2009}, the average particle-hole susceptibility
$\langle\chi_{\text{ph}}\rangle$ is added to $1/U$ or $m/4\pi a$. In other
works \cite{Heiselberg2000,KimTorma2009,TormaPethick2009}, only the
lowest order particle-hole diagram is considered so that their induced
interaction $U_{\text{ind}}^0 = -U^2 \langle\chi_{\text{ph}}^0\rangle $ is added to
$U$. Therefore, these works have to rely on the assumption $N(0)U \ll
1$ and the validity of the BCS mean-field result in order to obtain
the result of Eq.~(\ref{eq:gammaTc}). Away from the weak interaction
regime, a full summation of the particle-hole $T$ matrix becomes
necessary.

While the results from all different treatment seem to agree
quantitatively in the weak coupling limit, we expect to see
significant departures as the pairing interaction strength increases,
especially in the unitary regime.

With the overall effective interaction $U_\text{eff}$, the self
energy, as obtained from $\Sigma(K) = \sum_Q t_2(Q) G_0(Q-K)$, will
follow the same form as Eq.~(\ref{eq:Sigma}) although the gap values
will be different. Therefore, the fermion number equation will also
take the same form as Eq.~(\ref{eq:number}). Furthermore, the
pseudogap equation, given by $\Delta_\text{pg}^2 = -\sum_Q t_2(Q)$,
will also take the same form as Eq.~(\ref{eq:PG}).

Equations (\ref{eq:number}), (\ref{eq:PG}), and (\ref{eq:gap_ph}) now form a
new closed set, and will be solved to investigate the effect of the
particle-hole channel.

Note that in a \emph{very dilute} Fermi gas shifting $m/4\pi a$ by
$\langle \chi_\text{ph} \rangle $ has no significant influence in
experimental measurement of the $s$-wave scattering length $a$,
because $\langle \chi_\text{ph} \rangle $ has dimension
$[k_{\text{F}}]^3/[E_{\text{F}}] = [k_{\text{F}}]$ and thus vanishes
as $k_{\text{F}} \rightarrow 0$ in the zero density limit. However, a
finite $k_{\text{F}}$ will indeed shift the resonance location except
at very high $T$ where $\mu < 0$. In Ref.~\cite{Grimm5}, from which
the scattering lengths are often quoted for $^6$Li, the actual density
is comparable or even higher than that in most typical Fermi gas
experiments. Therefore, the particle-hole channel may play an
important role.

\emph{Here we propose that this particle-hole channel effect may be
  verified experimentally} by precision measurement of the magnetic
field $B$ at the exact Feshbach resonance point as a function of
density or $k_{\text{F}}$ at low $T$. The zero density field $B_0$ can
be obtained by extrapolation. Then one should have the field detuning
$\delta B = B-B_0 \propto k_{\text{F}}$.  Because different theories
predict a very different value of $\langle \chi_\text{ph} \rangle $ at
unitarity, the measured field detuning can thus be used to quantify
$\langle \chi_\text{ph}\rangle $ and test these theories. In
principle, one may experimentally measure $\langle \chi_\text{ph}
\rangle $ through the entire BCS-BEC crossover. For a Fermi gas in a
trap, the trap inhomogeneity leads to a distribution of
$k_{\text{F}}$. Instead of a uniform shift, this inhomogeneity will
spread out the unitary point at zero density into a narrow band at
finite density. The band width and mean shift are both expected to be
proportional to $k_{\text{F}}$. Such effect deserves further
investigation.

\section{Numerical Results and Discussions}
\label{sec:Results}

\subsection{Effect of particle-hole channel on BCS-BEC crossover}
\label{subsec:Results}

In this section, we will investigate the effect of the particle-hole
channel on the BCS-BEC crossover behavior, in terms of zero
temperature gap $\Delta(0)$, $T_{\text{c}}$ and their ratio. 

\begin{figure}[tb]
\centerline{\includegraphics[width=3.2in,clip]{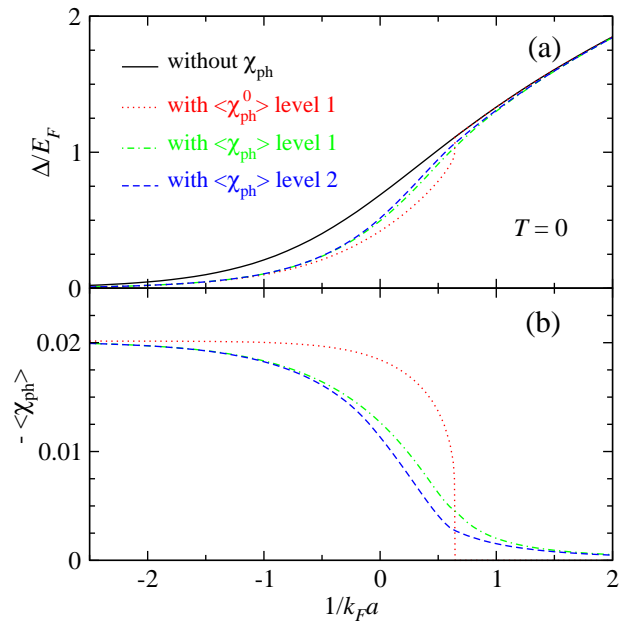}}
\caption{(Color online) Effect of the particle-hole channel
  contributions on the zero temperature gap in BCS-BEC crossover. In
  (a), the black solid curve is the gap without the particle-hole
  effect. The rest curves are calculated with the particle-hole
  channel effect but at different levels, i.e., using undressed
  particle-hole susceptibility $\langle \chi_{\text{ph}}^0\rangle $ with
  level 1 averaging (red dotted line), dressed particle-hole
  susceptibility $\langle \chi_{\text{ph}}\rangle $ with level 1 (green
  dot-dashed curve) and level 2 (blue dashed line) averaging,
  respectively. The corresponding values of the average particle-hole
  susceptibility with a minus sign are plotted in (b), in units of
  $k_{\text{F}}^3/E_{\text{F}}$. The particle-hole channel effect can be essentially
  neglected beyond $1/k_{\text{F}}a > 1.5$.}
\label{fig:zeroTgap}
\end{figure}

First, in Fig.~\ref{fig:zeroTgap}, we show the effect on the zero $T$
gap by comparing the calculated result with and without the
particle-hole channel contributions. Shown respectively in panel (a)
and (b) are plots of the zero $T$ gap $\Delta$ and the corresponding
particle-hole susceptibility (with a minus sign) as a function of
$1/k_{\text{F}}a$. The black solid line in Fig.~\ref{fig:zeroTgap}(a)
is the result without the particle-hole channel effect, whereas the
other curves are calculated with the effect at different levels of
approximation. The (red) dotted curve are calculated using the
undressed susceptibility $\langle \chi_{\text{ph}}^0\rangle $ at
average level 1. The (green) dot-dashed and (blue) dashed curves are
calculated using the dressed particle-hole susceptibility $\langle
\chi_{\text{ph}}\rangle $ with level 1 (green dot-dashed curve) and
level 2 (blue dashed line) averaging, respectively. The level 2 result
shows a slightly weaker particle-hole channel effect, as can be
expected from Fig.~\ref{fig:chi_ph_avg-k}.

One feature that is easy to spot is that the undressed particle-hole
susceptibility $\langle \chi_{\text{ph}}^0\rangle $ has a very abrupt
shut-off where the chemical potential $\mu$ changes sign. As a result,
the corresponding (red dotted) curve of the gap 
also merges abruptly with the (black solid) gap curve calculated
without particle-hole channel effect.  This is \emph{not unexpected}
as one can see from Eq.~(\ref{eq:Rechi0_nu=0_2}) that $\langle
\chi_{\text{ph}}^0\rangle =0$ at $T=0$ for $\mu\le 0$.  Furthermore,
Eq.~(\ref{eq:Rechi0_nu=0}) implies that $\langle
\chi_{\text{ph}}^0\rangle $ approaches zero at $\mu=0$ abruptly with a
finite slope as $k_\mu$ does. In contrast, with the self-energy
feedback included, either level 1 (green dot-dashed curves) or level 2
(blue dashed curves) average of $\langle \chi_{\text{ph}}\rangle $
approaches 0 smoothly as the BEC regime is reached. Consequently, in
Fig.~\ref{fig:zeroTgap}(a), the (green) dot-dashed and (blue) dashed
curves approach the (black) solid curve very gradually. It is also
worth pointing out that the difference between level 1 and level 2
average of $\langle \chi_{\text{ph}}\rangle $ is less dramatic than
that between $\langle \chi_{\text{ph}}\rangle $ and the undressed
$\langle \chi_{\text{ph}}^0\rangle $. Indeed, the (green) dot-dashed
and (blue) dashed curves are very close to each other.

The abrupt shut-off of $\langle \chi_{\text{ph}}^0\rangle $ at $\mu=0$ is
determined by the step function characteristic of the Fermi function
at $T=0$. At finite $T$, this shut-off will become smoother with an
exponential tail on the BEC side.

In the unitary regime, especially for $1/k_{\text{F}}a \in [-0.5, +0.5]$, the
particle-hole susceptibility is strongly over-estimated by the
undressed $\langle \chi_{\text{ph}}^0\rangle $ in comparison with the dressed
$\langle \chi_{\text{ph}}\rangle $. In this regime, both the gap and the
underlying Fermi surface (as defined by the chemical potential) are
large, so that neglecting the self-energy feedback leads to a strong
over-estimate of $\langle \chi_{\text{ph}}^0\rangle $, because the large gap
serves to suppress particle-hole fluctuations.

From Fig.~\ref{fig:zeroTgap}, we conclude that the particle-hole
effect diminishes quickly as the Fermi gas is tuned into the BEC
regime with increasing pairing interaction strength. Beyond
$1/k_{\text{F}}a>1.5$, the effect can essentially be neglected. For the level 1
average of the undressed particle-hole susceptibility, $\langle
\chi_{\text{ph}}^0\rangle $, as has been done in the literature, this effect
disappears immediately once the BEC regime (defined by $\mu<0$) is
reached, as far as the zero $T$ gap is concerned.

As a consistency check, we notice that in the BCS limit, the average
particle-hole susceptibility in all cases in
Fig.~\ref{fig:zeroTgap}(b) approaches the same asymptote, which is
given by Eq.~(\ref{eq:Rechi0_avg}). This confirms our previous
analytical analysis.

\begin{figure}[tb]
\centerline{\includegraphics[width=3.2in,clip]{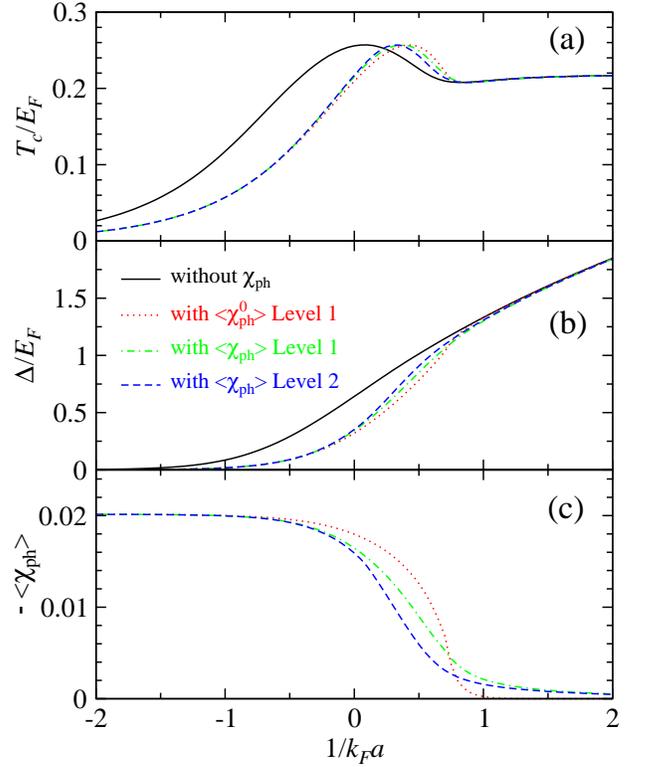}}
\caption{(Color online) Effect of the particle-hole channel
  contributions on $T_{\text{c}}$ and the pseudogap $\Delta$ at $T_{\text{c}}$ in
  BCS-BEC crossover. In (a) and (b), the black solid curves are
  calculated without the particle-hole effect. The rest curves are
  calculated with the particle-hole channel effect but at different
  levels, using undressed particle-hole susceptibility $\langle
  \chi_{\text{ph}}^0\rangle $ with level 1 averaging (red dotted line),
  dressed particle-hole susceptibility $\langle \chi_{\text{ph}}\rangle
  $ with level 1 (green dot-dashed curve) and level 2 (blue dashed
  line) averaging, respectively. The corresponding values of the
  average particle-hole susceptibility with a minus sign are plotted
  in (c), in units of $k_{\text{F}}^3/E_{\text{F}}$. The particle-hole channel effect
  can be essentially neglected beyond $1/k_{\text{F}}a > 1.5$.}
\label{fig:Tc}
\end{figure}

Next, we show in Fig.~\ref{fig:Tc} the effect of the particle-hole
channel on the behavior of $T_{\text{c}}$ as well as the pseudogap at
$T_{\text{c}}$.  Figure \ref{fig:Tc}(c) can be compared with
Fig.~\ref{fig:zeroTgap}(b). The curves for levels 1 and 2 average of
$\langle \chi_{\text{ph}} \rangle$ in Fig.~\ref{fig:Tc}(c) are very
similar to those in Fig.~\ref{fig:zeroTgap}(b), with the values at
$1/k_{\text{F}}a=0$ slightly smaller. On the other hand, the curve for
$\langle \chi_{\text{ph}}^0 \rangle$ has a smooth thermal exponential 
tail in the BEC regime in Fig.~\ref{fig:Tc}(c). 
Thus, the pseudogap $\Delta(T_{\text{c}})$ calculated using
$\langle \chi_{\text{ph}}^0 \rangle$ now merges back to the (black)
solid curve smoothly.

Similar to the zero $T$ gap case in Fig.~\ref{fig:zeroTgap}, the
difference in the effect on $T_{\text{c}}$ and $\Delta(T_{\text{c}})$ between level 1
and level 2 averaging mainly resides in the unitary regime, and is
less dramatic than that between undressed and dressed particle-hole
susceptibility. Again, \emph{the undressed particle-hole
  susceptibility gives rise to an overestimate of the particle-hole
  channel effect.}

In all cases, the particle-hole susceptibility becomes negligible in
the BEC regime. The effect of the particle-hole channel shifts the
$T_{\text{c}}$ and $\Delta(T_{\text{c}})$ curve towards larger
$1/k_{\text{F}}a$, although the amount of shift clearly depends on the
value of $1/k_{\text{F}}a$.

Now we study the effect of the particle-hole channel on the ratio
$2\Delta(0)/T_{\text{c}}^{}$. It suffices to consider the mean-field
ratio, $2\Delta(0)/T_{\text{c}}^{\text{MF}}$, since
$2\Delta(0)/T_{\text{c}}^{}$ obviously will deviate from the weak
coupling BCS result when pairing fluctuations are included in the
crossover and BEC regimes. From Fig.~\ref{fig:chi_ph_avg-k}, we see a
strong $T$ dependence of the particle-hole susceptibility. Therefore,
the effect on $T_{\text{c}}^{\text{MF}}$ and on zero $T$ gap
$\Delta(0)$ are different, as can be seen roughly from
Figs.~\ref{fig:zeroTgap} and \ref{fig:Tc}.

\begin{figure}[tb]
\centerline{\includegraphics[width=3.2in,clip]{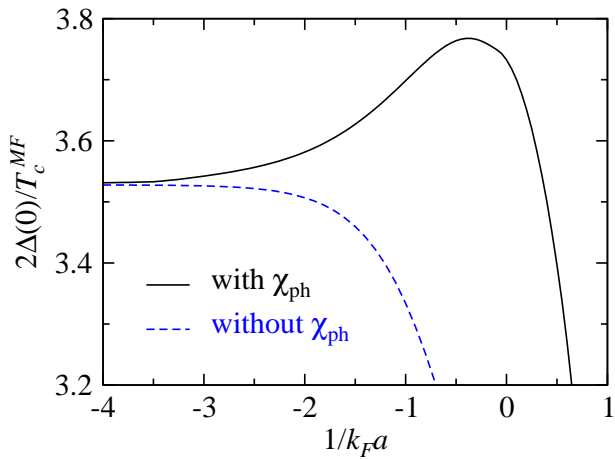}}
\caption{(Color online) Effect of the particle-hole channel
  contributions on the ratio $2\Delta(0)/T_{\text{c}}^{\text{MF}}$ in BCS-BEC
  crossover. Shown is the mean-field ratio calculated with (black
  solid curve) and without (blue dashed curve) the particle-hole
  channel contributions.  Here the particle-hole susceptibility
  $\langle \chi_{\text{ph}}\rangle $ is calculated with level 2 averaging.}
\label{fig:ratio}
\end{figure}

In Fig.~\ref{fig:ratio}, we plot this mean-field ratio as a function
of $1/k_{\text{F}}a$ with (black solid curve) and without (blue dashed
curve) the particle-hole channel effect. Here the particle-hole
susceptibility $\langle \chi_{\text{ph}}\rangle $ is calculated with
level 2 averaging. In the $1/k_{\text{F}}a \rightarrow -\infty$ limit,
the ratio is unaffected by the particle-hole channel. As
$1/k_{\text{F}}a$ increases, the contribution of the particle-hole
channel causes this ratio to increase gradually. At $1/k_{\text{F}}a =
-4$, which is still a very weak pairing case, the ratio is already
slightly larger. The effect is most dramatic in the unitary regime,
since further into the BEC regime, $\langle \chi_{\text{ph}}\rangle $
will vanish gradually. It is worth noting that even without the
particle-hole channel, the ratio $2\Delta(0)/T_{\text{c}}^{\text{MF}}$
starts to decrease from its weak coupling limit, $2\pi \e^{-\gamma}
\approx 3.53$.

Finally, we estimate the shift in Feshbach resonance positions.  From
Figs.~\ref{fig:zeroTgap} and \ref{fig:Tc}, we find that
$\chi_\text{ph}$ does not necessarily diminish as $T$ increases except
at very high $T$ (where $\mu$ becomes negative, so that
$|\chi_\text{ph}|$ will decrease exponentially with $T$.) In fact,
this can be understood because $\Delta(T)$ decreases with $T$ so that
$|\chi_\text{ph}|$ increases.  We take $\langle \chi_\text{ph}\rangle
= - 0.01 k_\text{F}^3/E_\text{F} = -0.01 (2m
k_\text{F}/\hbar^2)$. According to Eq.~(\ref{eq:gap_ph}), the shift in
$1/a$ is $\delta (1/a) = -4\pi \hbar^2 \langle \chi_\text{ph}\rangle
/2m = 0.08 \pi k_\text{F}$. In other words, the dimensionless shift
$\delta (1/k_\text{F}a) = 0.25$, which is independent of density and
is no longer negligible. This is in good agreement with the actual
shift $0.32$ of the peak location of the $T_\text{c}$ curve in
Fig.~\ref{fig:Tc}(a). For a typical $T_\text{F} = 1\mu$K in $^6$Li,
using the approximate expression $a = a_\text{bg} [1-W/(B-B_0)]$, we
obtain the shift in resonance position $\delta B_0 = - 0.08 W
(k_\text{F}a_\text{bg}) = 7.8$G. Here for the lowest two hyperfine
states, the resonance position $B_0 = 834.15$G, the resonance width
$W=300$G, and the background scattering length $a_\text{bg} = -1405
a_0$, with $a_0 = 0.528$\AA. Clearly, the shift $\delta B_0$ is not
small. In reality, one needs to solve self-consistently the equation
$m/(4\pi a) + \langle \chi_\text{ph}\rangle = 0$, and take care of the
trap inhomogeneity. These will likely make the actual average shift
smaller.

The susceptibility $\chi_\text{ph}$ calculated with and without the
self energy feedback differs by roughly a factor of 2 at
unitarity. This can be used to test different theories, as mentioned
earlier.

A question arises naturally as to whether the particle-hole channel
effect has already been included in the experimentally measured
scattering length $a$, since, after all, the measurements of $a$ such
as those in Ref.~\cite{Grimm5} were carried out at densities
comparable to typical Fermi gas experiments. This also depends on
whether the temperature was high enough during the measurements.

\subsection{Critical temperature $T_{\text{c}}$ at unitarity}
\label{subsec:Tc}

Finally, we compare our result on the critical superfluid transition
temperature $T_{\text{c}}/E_{\text{F}}$ for a 3D homogeneous Fermi gas
at unitarity
with those reported in the literature. From Fig.~\ref{fig:Tc}, we read
$T_{\text{c}}/E_{\text{F}}=0.217$ using level 2 average of
$\langle\chi_{\text{ph}}\rangle$. And the maximum $T_{\text{c}}\approx
0.257 $ now occurs at $1/k_{\text{F}}a \approx 0.32$, on the BEC
side. The level 1 average of $\langle\chi_{\text{ph}}^0\rangle$ yields
a slightly lower value, $T_{\text{c}}/E_{\text{F}}=0.209$. However, we
emphasize that the level 2 average of $\langle\chi_{\text{ph}}\rangle$
is more reasonable. Note that as in the theory without particle-hole
channel effect, we have dropped out the incoherent part of the
self-energy from particle-particle scattering. Inclusion of the
incoherent part is necessary in order to obtain the correct value of
the $\beta$ factor.

Hu and coauthors \cite{Drummond3,Drummond5} have been claiming to be
able to obtain the correct value of the $\beta$ factor, using an
NSR-based approach, \emph{without including the particle-hole
  channel}. Obviously, their claim will breakdown when the
particle-hole channel is included.

The value of $T_{\text{c}}$ for a homogeneous Fermi gas at unitarity
has been under intensive study over the past few years. Without the
particle-channel effect, using the pairing fluctuation theory, we
previously reported a value of $T_{\text{c}}/E_{\text{F}}= 0.255$ for
a short range Gaussian potential (using a two-channel model with a
cutoff momentum $k_0/k_{\text{F}}=80$) \cite{Reviews2} or 0.256 for an
exact contact potential. In all cases, the maximum $T_{\text{c}}$
occurs very close to but slightly on the BEC side of the
unitarity. The original NSR theory \cite{NSR} predicted
$T_{\text{c}}/E_{\text{F}}= 0.222$.  Haussmann \textit{et al.}
\cite{Zwerger} found 0.16, using a conserving approximation which
involves only dressed Green's functions. However, this theory did not
contain a pseudogap at $T_{\text{c}}$ and exhibited unphysical
non-monotonic first order like behavior in the temperature dependence
of entropy $S(T)$. Floerchinger \textit{et al.} \cite{Floerchinger}
found a high value of 0.264 even after including the particle-hole
channel fluctuations. By including the induced interaction in an
NSR-based treatment without self-energy feedback, Yin and coworkers
\cite{Yin2009} reported a value of 0.178. However, as shown above, the
neglect of the pseudogap self energy feedback in Ref.~\cite{Yin2009}
has caused a serious over-estimate of the contributions of the induced
interaction. In addition, their work suffers all defects of the NSR
theory, i.e., lack of self-consistency and neglect of the pseudogap
effect.

Using the renormalization group method, Stoof and coworkers
\cite{StoofPRL2008} obtained $T_{\text{c}}/E_{\text{F}} = 0.13$, lower
than most other calculations. Troyer and coworkers
\cite{TroyerPRL2006} reported 0.152 using quantum Monte Carlo (QMC)
simulations for lattice fermions at finite densities and then
extrapolated to zero density limit. 
Using (and improved upon) the method of Ref.~\cite{TroyerPRL2006},
Goulko and Wingate \cite{Wingate} found a higher value, 0.171. An even
higher value would have been obtained had they used an quadratic fit
to (and extrapolation of) their intermediate density data. Also using
QMC, Akkineni \textit{et al.}  \cite{Akkineni} recently reported a
value of 0.245.

Experimentally, only $T_{\text{c}}$ in a trap has been measured. The
Duke group \cite{ThermoScience}, in collaboration with Chen \textit{et
  al.}, found $T_{\text{c}}/E_{\text{F}} = 0.27$ in a trap through a
thermodynamic measurement in a unitary $^6$Li gas. Later, the Duke
group \cite{Thomas2007,ThomasJLTP} obtained 0.29 and 0.21 by fitting
entropy and specific heat data with different formulas. At unitarity,
it has been known that $T_{\text{c}}/E_{\text{F}}$ in the trap is only
slightly higher than its homogeneous counterpart.  Therefore, these
experimental values probably imply that the homogeneous value of
$T_{\text{c}}/E_{\text{F}}$ is about 0.25 $\sim$ 0.19. Our result,
$T_{\text{c}}/E_{\text{F}} =0.217$, is in reasonable agreement with
these experiments. Recently, Ku \textit{et al.} \cite{Zwierlein2011}
reported $T_\text{c}/E_\text{F} \approx 0.167$ for a homogeneous Fermi
gas by identifying the lambda-like transition
temperature. Interestingly, if we take an constant $\delta\Sigma =
-0.3E_\text{F}$ (half of the energy of a single spin down atom in a
spin up Fermi sea \cite{Combescot}) as the incoherent part of the self
energy in Eq.~(\ref{eq:Sigma_pg}), $T_\text{c}/E_\text{F}$ will be
suppressed down to 0.174, close to this recent experimental
value. Full numerical inclusion of $\delta\Sigma$ will be done in a
future study.

\begin{figure}
\centerline{\includegraphics[width=3.2in,clip]{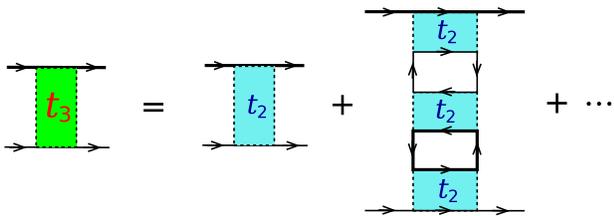}}
\label{fig:t3}
\caption{(Color online) Higher order $T$-matrix, $t_3$, obtained by
  repeating the $T$-matrix $t_2$.}
\end{figure}

\subsection{Higher order corrections}

In addition to non-ladder diagrams, which we have chosen not to
consider, there seem to be a series of higher order corrections. For
example, one can imagine repeating the $T$-matrix $t_2$ in the way
shown in Fig.~\ref{fig:t3}, and obtaining a higher order $T$-matrix
$t_3$. Such $t_3$ can then be repeated to obtain a higher order
$T$-matrix $t_4$, and so on. While one may argue these higher order
$T$-matrices are indeed of higher order in bare interaction $U$, our
experience with $t_2$ seems to imply that detailed study needs to be
carried out before we jump to a conclusion on this. Indeed, even the
lowest order so-called induced interaction $U_{\text{ind}}^0$ is one order
higher in $U$ than $U$ itself.

\section{Conclusions}
\label{sec:conclusions}

In summary, we have studied the effects of the particle-hole channel
on BCS-BEC crossover and compared with lower level approximations.  We
include the self-energy feedback in the particle-hole susceptibility
$\chi_{\text{ph}}$, which leads to substantial differences than the result
without self-energy feedback. 

We have investigated the dynamic structure of $\chi_{\text{ph}}$, and have
discovered very strong temperature, momentum and frequency
dependencies. Angular (as well as radial) average in the momentum
space of the particle-hole susceptibility has been done in order to
keep the equations manageable. We have performed the average at two
different levels and also compared with the result calculated without
including the self-energy feedback. We conclude that the level 2
averaging, i.e., both over angles and a range of momentum, is more
reasonable. Computations of the particle-hole susceptibility without
the self-energy feedback leads to an overestimate of the particle-hole
channel effect.

In the weak coupling BCS limit, our result agrees, to the leading
order, with that of GMB and others in the literature. Away from the
weak coupling limit, $\Delta(0)$ and $T_{\text{c}}$ are suppressed
differently. We have also studied the ratio
$2\Delta(0)/T_{\text{c}}^{\text{MF}}$ at the mean-field level and
found that it is modified by the particle-hole fluctuations.  The
particle-hole channel effects diminish quickly once the system enters
BEC regime.

Without including the incoherent part of the self energy from
particle-particle scattering, our present result on the critical
temperature at unitarity yields $T_{\text{c}}/E_{\text{F}}\approx
0.217$, substantially lower than that obtained without the
particle-hole effect. This value agrees reasonably well with some
existing experimental measurement.

We have also made a falsifiable proposal that the particle-hole
contribution can be measured by locating the Feshbach resonance
positions as a function of $k_\text{F}$ and that this can be used to
test different theories.

To study more accurately the quantitative consequences of the dynamic
structure of the particle-hole susceptibility, full-fledged numerical
calculations are needed, without taking simple angular average and
setting frequency $\nu=0$.  Further investigation is called for in
order to determine whether higher order $T$-matrices will make a
significant difference or not.

\acknowledgments

This work is supported by NSF of China (grant No. 10974173), MOST of
China (grant Nos. 2011CB921300 and 2011CC026897), the Fundamental
Research Funds for the Central Universities of China (Program
No. 2010QNA3026), Changjiang Scholars Program of the Ministry of
Education of China, Qianjiang RenCai Program of Zhejiang Province
(No. 2011R10052), and by Zhejiang University (grant No. 2009QNA3015).


\appendix

\begin{figure}[tb]
\centerline{\includegraphics[width=3.2in,clip]{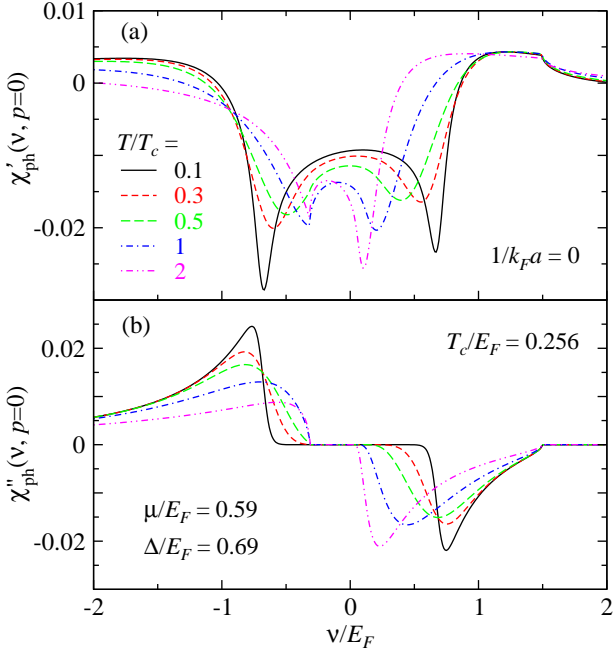}}
\caption{(Color online) The particle-hole susceptibility
  $\chi_{\text{ph}}^R(\nu,0)$ at zero momentum $p$ in the presence of
  self-energy feedback for different $T$ (as labeled) at unitarity. To
  single out the temperature effect, we fix $\Delta=0.686$ and
  $\mu=0.59$ at their values at $T=0$, calculated using the pairing
  fluctuation theory without the particle-hole channel effect.}
\label{fig:chi_ph-T}
\end{figure}

\begin{figure}[tb]
\centerline{\includegraphics[width=3.2in,clip]{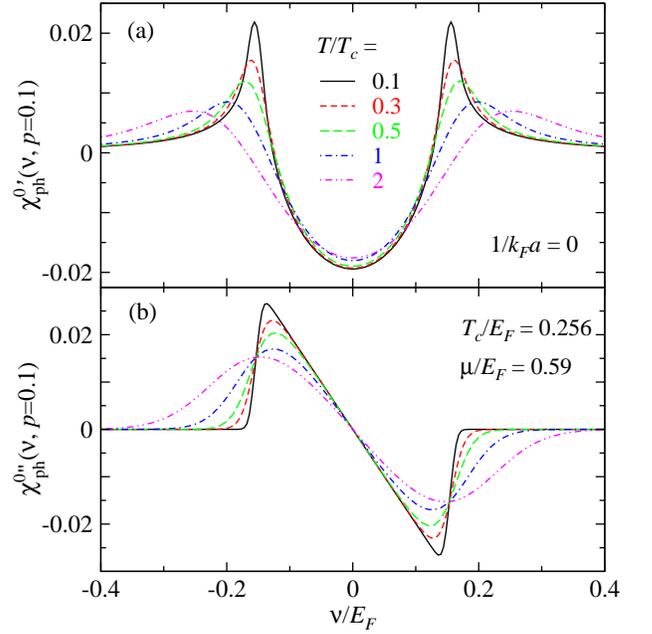}}
\caption{(Color online) The undressed particle-hole susceptibility
  $\chi_{\text{ph}}^{0R}(\nu,p=0.1)$ for different $T$ (as labeled) at
  unitarity and momentum $p=0.1$. As in Fig.~\ref{fig:chi_ph-T}, we
  fix $\mu=0.59$.}
\label{fig:chi0_ph-T}
\end{figure}

\section{Dynamic structure of the particle-hole susceptibility} 

In this Appendix, we will present a series of two-dimensional plots, in order
to make the 3D data shown in Fig.~\ref{fig:chi_ph} quantitatively
easier to read.

\begin{figure}[tb]
\centerline{\includegraphics[width=3.2in,clip]{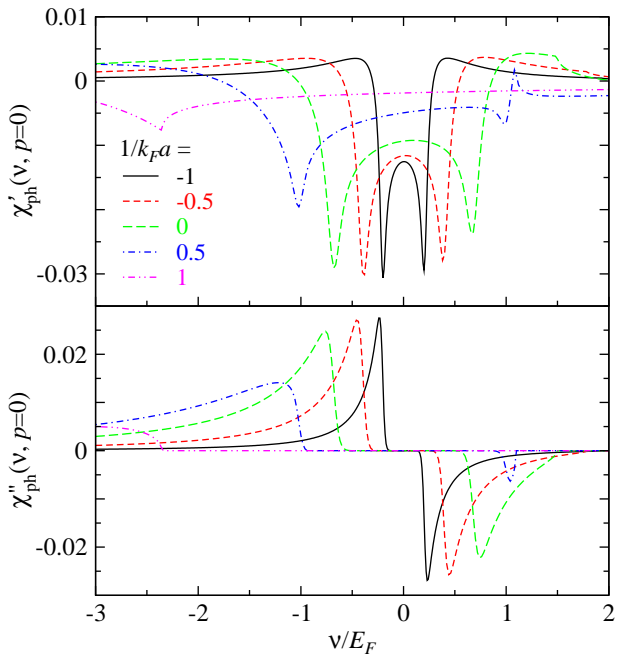}}
\caption{(Color online) The real and imaginary parts of particle-hole
  susceptibility $\chi_{\text{ph}}^R(\nu,0)$ in the presence of self-energy
  feedback for various values of $1/k_{\text{F}}a$ from BCS to BEC. The curves
  are calculated at $0.1T_{\text{c}}$. For each case, the parameters $\Delta$,
  $\mu$ and $T_{\text{c}}$ are calculated using the pairing fluctuation theory
  without the particle-hole channel effect.}
\label{fig:chi_ph-InvkFa}
\end{figure}

First, we study the impact of temperature. We present in
Fig.~\ref{fig:chi_ph-T} the real and imaginary parts of
$\chi_{\text{ph}}^R(\nu, 0)$ for different $T$ from low to high in the
unitary limit, $1/k_{\text{F}}a=0$. To single out the temperature effect, here
we take for all temperature $\Delta=0.686$ and $\mu=0.59$, which are
their values calculated at $T=0$ using the pairing fluctuation theory
without the particle-hole channel effect. Evidently, beyond the point
$\nu = \sqrt{\mu^2+\Delta^2}+\mu =1.49$, the imaginary part
$\chi_{\text{ph}}^{\prime\prime}(\nu, 0)$ vanishes identically.  Around
$\nu=0$, the lower bound of the range of $\nu$ where
$\chi_{\text{ph}}^{\prime\prime}(\nu, 0)$ essentially vanishes changes from
$\nu = -\Delta = -0.686$ at low $T$ to $\nu =
-(\sqrt{\mu^2+\Delta^2}-\mu)=0.315$ at high $T$. Meanwhile, its upper
bound decreases continuously with $T$ from $\nu =\Delta$ at low $T$ to
$\nu = 0$ at very high $T$. This numerical result agrees with our
previous analysis. A comparison with the real part reveals that the
peaks in $\chi_{\text{ph}}^{\prime}(\nu, 0)$ correspond to the sharp rises in
the plot of $\chi_{\text{ph}}^{\prime\prime}(\nu, 0)$ near these lower and
upper bounds. This can also be seen from the Kramers-Kronig relation
between the real and imaginary parts of $\chi_{\text{ph}}^R(\nu,0)$.

\begin{figure}[tb]
\centerline{\includegraphics[width=3.3in,clip]{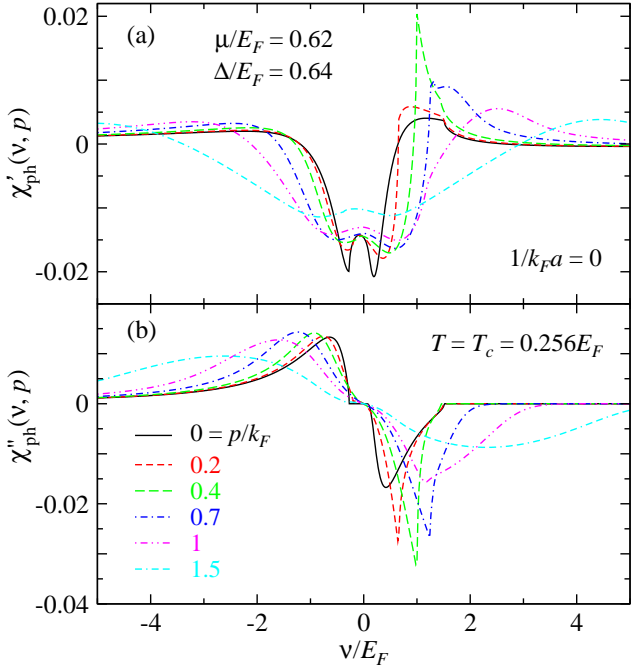}}
\caption{(Color online) The particle-hole pair susceptibility
  $\chi_{\text{ph}}^R(\nu,p)$ at unitarity and at $T_{\text{c}}$ in the presence of
  self-energy feedback for increasing $p=0$, 0.2, 0.4, 0.7, 1, and
  1.5. The parameters $\Delta$, $\mu$ and $T_{\text{c}}$ are calculated using
  the pairing fluctuation theory without the particle-hole channel
  effect. As $p$ increases, the peaks becomes broader and smeared out,
  and the upper bound in $\nu$ beyond which
  $\chi_{\text{ph}}^{\prime\prime}(\nu,p)$ vanishes increases towards
  infinity. In addition, the well defined gap at $p=0$ in
  $\chi_{\text{ph}}^{\prime\prime}(\nu,p)$ near $\nu=0$ gradually
  disappears.}
\label{fig:chi_ph-p}
\end{figure}

In comparison, we have also studied the temperature evolution of the
undressed $\chi_{\text{ph}}^{0R}(\nu,p)$.  Shown in Fig.~\ref{fig:chi0_ph-T}
is the result for $p=0.1$. From Fig.~\ref{fig:chi_ph}, it is easy to
see that one cannot plot the result for $p=0$. This can also be seen
from Eqs.~(\ref{eq:Rechi0_p=0}) and (\ref{eq:Rechi0_nu=0}). For finite
$p$, say $p=0.1$, the peaks at low $T$ in both real and imaginary
parts become more smeared out as $T$ increases. Near $\nu=0$, we see
that $\chi_{\text{ph}}^{0\prime\prime}(\nu,p=0.1)$ is proportional to $\nu$,
in agreement with our previous analysis. $\chi_{\text{ph}}^{0R}(\nu,p)$ shows
good symmetry about $\nu$: $\chi_{\text{ph}}^{0R}(-\nu,p) =
\chi_{\text{ph}}^{0R*}(\nu,p)$. There is no gap effects, of course.

Shown in Fig.~\ref{fig:chi_ph-InvkFa} is the evolution of the
particle-hole susceptibility $\chi_{\text{ph}}^{R}(\nu,0)$ at total momentum
$p=0$ in the presence of feedback effect with increasing pairing
strength. These curves are calculated at low temperature
$T=0.1T_{\text{c}}$. Here for each interaction strength, the parameters
$\Delta$, $\mu$ and $T_{\text{c}}$ are calculated using the pairing fluctuation
theory without the particle-hole channel effect. Around $\nu=0$, the
range within which the imaginary part vanishes is given by
$|\nu|<\Delta$ for the $\mu>0$ cases ($1/k_{\text{F}}a = -1$ through 0.5). For
$1/k_{\text{F}}a=1$, $\mu/E_{\text{F}}=-0.8$, the lower bound is given by
$-(\sqrt{\mu^2+\Delta^2} -\mu)=-2.35$ and its upper bound extends to
$\infty$ since $\Delta=1.33 > \sqrt{\mu^2+\Delta^2} +\mu = 0.75$. It is
obvious that this range becomes wider and wider with increasing
pairing strength from BCS to BEC.

From Figs.~\ref{fig:chi_ph}(a) and \ref{fig:chi_ph}(b), one readily
notice that for the undressed $\chi_{\text{ph}}^{0R}(\nu,p)$, it is not
appropriate to plot $\chi_{\text{ph}}^{0R}(\nu,p=0)$ as a function of
$\nu$. Instead, one has to plot at a finite $p$, say, $p=0.1k_{\text{F}}$, in
order to study its temperature evolution. Our result (not shown)
demonstrates that the real part presents a double peak structure, with
the peaks becoming increasingly broader as $T$ increases. At low $T$,
the location of the peaks are roughly given by $\nu = \pm pk_\mu/m
\approx \pm 0.15$ for $\mu = 0.59$ at unitarity. This relation also
shows how the $\chi_{\text{ph}}^{0R}(\nu,p)$ curves evolve with total
momentum $p$.

Next, we investigate how the particle-hole susceptibility
$\chi_{\text{ph}}^R(\nu,p)$ evolves with total momentum $p$ in the presence
of feedback effect. Shown in Fig.~\ref{fig:chi_ph-p} are the curves of
the real and imaginary parts for increasing $p$ for a unitary Fermi
gas, calculated at $T_{\text{c}}$. Just as in Fig.~\ref{fig:chi_ph-T}, the
$p=0$ curve shows a clear gap in the neighborhood of $\nu=0$ in the
imaginary part, $\chi_{\text{ph}}^{\prime\prime}(\nu,p)$. As $p$ increases,
this gap gradually disappears, and the upper bound in $\nu$ beyond
which $\chi_{\text{ph}}^{\prime\prime}(\nu,p)$ vanishes increases towards
infinity.  At the same time, the peaks in the real part becomes
broader and smeared out. From Fig.~\ref{fig:chi_ph-p}(a), we can see that
at $\nu=0$, the real part slowly increases with $p$.

The zero frequency value $\chi_{\text{ph}}^\prime(0,p)$ is plotted in
Fig.~\ref{fig:chi_ph_nu0-p} in the text as a function of $p$.

\begin{figure}[b]
\centerline{\includegraphics[width=3.2in,clip]{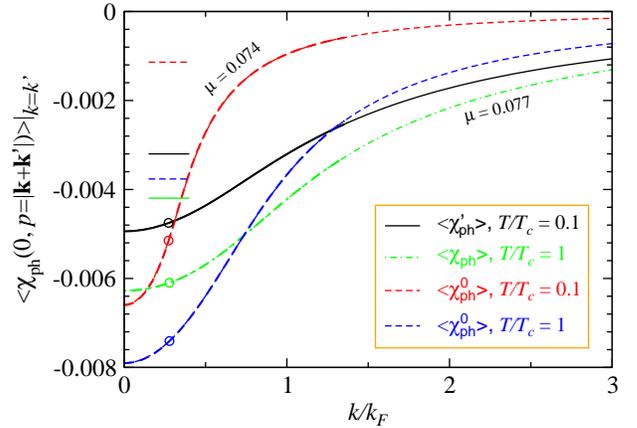}}
\caption{(Color online) Angular average of the on-shell particle-hole
  susceptibility, $\langle
  \chi_{\text{ph}}(0,p=|\mathbf{k}+\mathbf{k}'|)\rangle $ at $\nu=0$
  as a function of momentum $k/k_{\text{F}}$, under the condition
  $k=k'$, calculated at $1/k_{\text{F}}a=0.5$. The conventions and
  legends are the same as in Fig.~\ref{fig:chi_ph_avg-k}. Here
  $T_{\text{c}}=0.226E_{\text{F}}$ and the chemical potential
  $\mu/E_{\text{F}} = 0.077$ and 0.074 at $T_{\text{c}}$ and
  $0.1T_{\text{c}}$, respectively. Clearly, there are even stronger
  temperature and $k$ dependencies in both $\langle
  \chi_{\text{ph}}(0,p)\rangle $ and $\langle
  \chi_{\text{ph}}^0(0,p)\rangle $ than the unitary case shown in
  Fig.~\ref{fig:chi_ph_avg-k}. The (absolute) values of Level 2
  average are substantially smaller than their level 1 counterpart. }
\label{fig:chi_ph_avg-k_BEC}
\end{figure}

Finally, we show in Fig.~\ref{fig:chi_ph_avg-k_BEC} the angular
average of the on-shell particle-hole susceptibility, $\langle
\chi_{\text{ph}}(0,p=|\mathbf{k}+\mathbf{k}'|)\rangle $ at $\nu=0$ as
a function of momentum $k/k_{\text{F}}$, under the condition $k=k'$,
calculated at $1/k_{\text{F}}a=0.5$. The chemical potential is nearly
zero, close to the boundary separating fermionic and bosonic
regimes. In comparison with the unitary case shown in
Fig.~\ref{fig:chi_ph_avg-k}, we conclude that both dressed and
undressed particle-hole susceptibility exhibit stronger temperature
and $k$ dependence. Here the small chemical potential determines that
the susceptibility is also much smaller. It is worth mentioning that
the level 1 average of the undressed particle-hole susceptibility
actually shows a much stronger temperature dependence. This is because
$1/k_{\text{F}}a$ is very close to the fast shut-off shown in
Fig.~\ref{fig:zeroTgap}.

\bibliographystyle{apsrev4-1}

\begin{thebibliography}{92}
\bibitem [{Not()}]{NotesonG0Gph}%
  \BibitemOpen
  \href@noop {} {}\bibinfo {note} {The contribution of the particle-hole
  channel $T$-matrix appears (as the last diagram) in Fig.~2.2(a) of
  Ref.~\protect\onlinecite{ChenPhD}, but was later dropped in the derivation of
  the pairing fluctuation theory, which focuses primarily on the pairing
  phenomena in particle-particle channel. Nevertheless, one can see from this
  figure that the mixing of $G_0$ and $G$ appears in both the particle-particle
  and particle-hole susceptibilities.}\BibitemShut {Stop}%
\end{thebibliography}

%

\end{document}